\documentclass[12pt]{article}

\usepackage{times}

\usepackage{amssymb,amsfonts,amsmath,amsthm,epsfig,url}
\usepackage[numbers,sort&compress]{natbib}

\newcommand{\R}{\mathbb{R}}

\newcommand{\bu}{\mathbf{u}}
\newcommand{\bA}{\mathbf{A}}
\newcommand{\bb}{\mathbf{b}}
\newcommand{\0}{\mathbf{0}}
\newcommand{\bS}{\mathbf{S}}
\newcommand{\bv}{\mathbf{v}}
\newcommand{\ba}{\mathbf{a}}
\newcommand{\bc}{\mathbf{c}}
\newcommand{\mopt}{\ensuremath{M_\text{opt}}}
\newcommand{\bx}{\mathbf{x}}
\newcommand{\by}{\mathbf{y}}
\def\bs{\mathbf{s}}

\newtheorem{theorem}{Theorem}

\begin{document} 

\vspace{2cm}
\noindent{\Large\bf Spontaneous Reaction Silencing in Metabolic Optimization} 
\vspace{5mm}

\noindent
{\large Takashi Nishikawa,$^{1, 4}$ Natali Gulbahce,$^{2, 3}$ Adilson E. Motter$^{4, \ast}$}
\vspace{2mm}
\noindent

{\noindent$^{1}$\footnotesize Division of Mathematics and Computer Science, Clarkson University,
Potsdam, NY 13699, USA.\\[0mm]
 $^{2}$Department of Physics and Center for Complex Network Research,
 Northeastern University, Boston, MA 02115, USA.\\[0mm]
 $^{3}$Center for Cancer Systems Biology, Dana Farber Cancer Institute, Boston, MA 02115, USA.\\[0mm]
 $^{4}$Department of Physics and Astronomy and Northwestern Institute on Complex Systems, 
 Northwestern University, Evanston, IL 60208, USA.\\[0mm]
 $^\ast$Corresponding author. Department of Physics and Astronomy, Northwestern University,
  2145 Sheridan Road, Evanston, IL 60208, USA;
  Tel.: +1 847 491 4611; Fax: +1 847 491 9982; E-mail: \mbox{motter@northwestern.edu}\\[4mm]
}

\begin{center}{\bf Abstract}\end{center}
Metabolic reactions of single-cell organisms are routinely observed to become dispensable or even incapable of carrying activity under certain circumstances.  Yet, the mechanisms as well as the range of conditions and phenotypes associated with this behavior remain very poorly understood.  Here we predict computationally and analytically that any organism evolving to maximize growth rate, ATP production, or any other linear function of metabolic fluxes tends to significantly reduce the number of active metabolic reactions compared to typical non-optimal states.  The reduced number appears to be constant across the microbial species studied and just slightly larger than the minimum number required for the organism to grow at all.  We show that this {\em massive spontaneous reaction silencing} is triggered by the irreversibility of a large fraction of the metabolic reactions and propagates through the network as a cascade of inactivity. Our results help explain existing experimental data on intracellular flux measurements and the usage of latent pathways, shedding new light on microbial evolution, robustness, and versatility for the execution of specific biochemical tasks. In particular, the identification of optimal reaction activity provides rigorous ground for an intriguing knockout-based method recently proposed for the synthetic recovery of metabolic function.

\vspace{5mm}
\noindent

\clearpage

\section*{Author Summary}
Cellular growth and other integrated metabolic functions are manifestations of the coordinated interconversion of a large number of chemical compounds.  But what is the relation between such whole-cell behaviors and the activity pattern of the individual biochemical reactions? In this study, we have used flux balance-based methods and reconstructed networks of {\it Helicobacter pylori}, {\it Staphylococcus aureus}, {\it Escherichia coli} and {\it Saccharomyces cerevisiae} to show that a cell seeking to optimize a metabolic objective, such as growth, has a tendency to spontaneously inactivate a significant number of its metabolic reactions, while all such reactions are recruited for use in typical suboptimal states. The mechanisms governing this behavior not only provide insights into why numerous genes can often be disabled without affecting optimal growth, but also lay a foundation for the recently proposed synthetic rescue of metabolic function, in which the performance of suboptimally operating cells can be enhanced by disabling specific metabolic reactions. Our findings also offer explanation for another experimentally observed behavior, in which some inactive reactions are temporarily activated following a genetic or environmental perturbation. The latter is of utmost importance given that non-optimal and transient metabolic behaviors are arguably common in natural environments.

\textwidth 16cm
\oddsidemargin 0.2cm
\baselineskip 15pt

\section*{Introduction}

A fundamental problem in systems biology is to understand how living cells adjust the usage pattern of their components to respond and adapt to specific genetic, epigenetic, and environmental conditions.  
In complex metabolic networks of single-cell organisms, there is mounting evidence in the experimental~\cite{Baba:2006ng,Joyce12012006,Giaever:2002db,RosarioGil09012004,Hashimoto2005,K.Kobayashi04152003} and modeling~\cite{Burgard:2001fk,Burgard:2001lr,Mahadevan:2003fk,Reed:2004fk,Pal2006,Henry:2006lr,Henry:2007lr,Feist:2007lr} literature that a surprisingly small part of the network can carry all metabolic functions required for growth in a given environment, whereas the remaining part is potentially necessary only under alternative conditions~\cite{Hillenmeyer2008}.
The mechanisms governing this behavior are clearly important for understanding systemic properties of cellular metabolism, such as mutational robustness, but have not received full attention.  This is partly because current modeling approaches are mainly focused on predicting whole-cell phenotypic characteristics without resolving the underlying biochemical activity.
These approaches are typically based on optimization principles, and hence, by their nature, do not capture processes involving non-optimal states, such as the temporary activation of latent pathways {\em during} adaptive evolution towards an optimal state~\citep{Fong2005,Fong2006}.

\begin{figure}[ht]
\begin{center}
\epsfig{figure=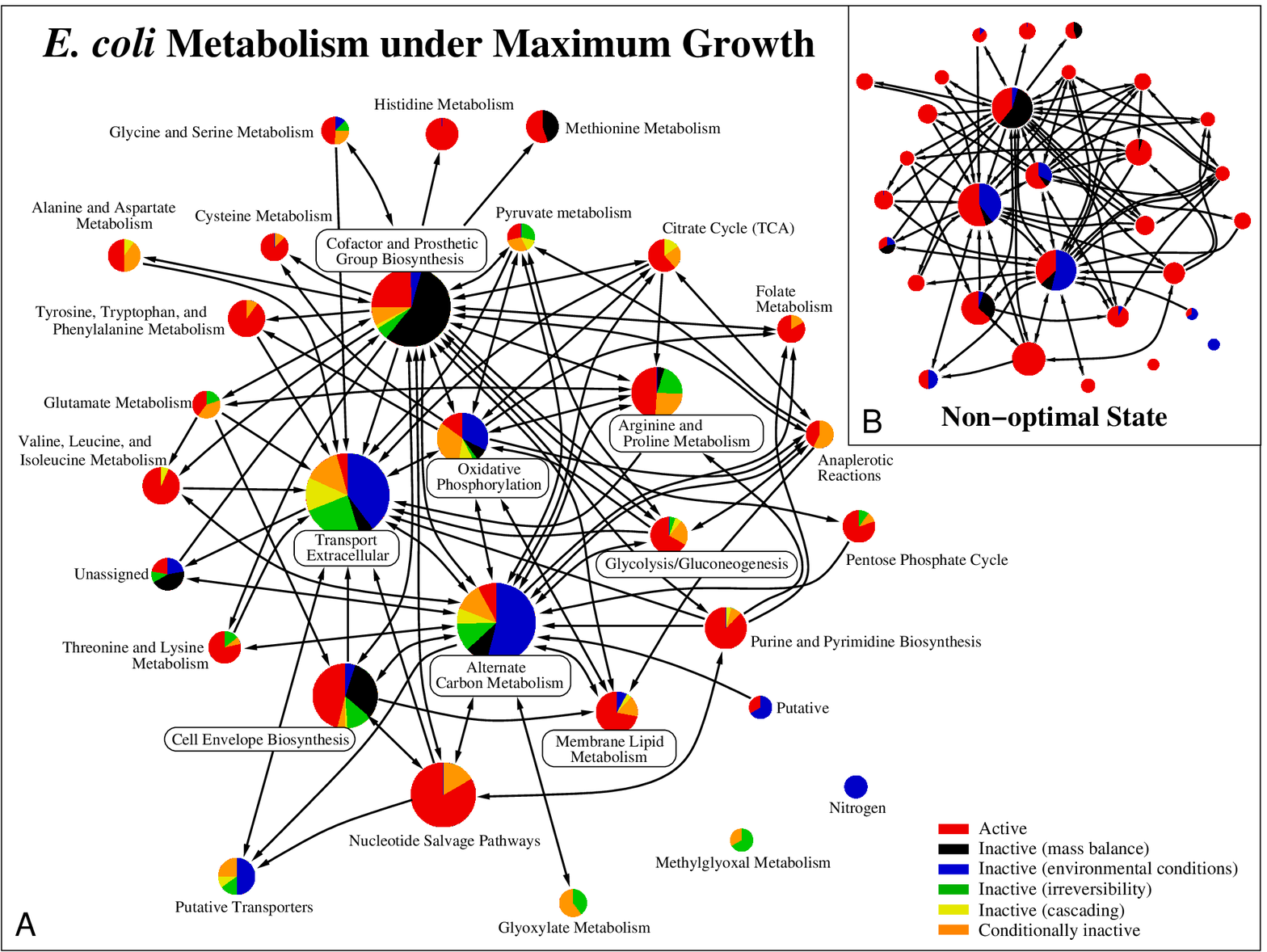,height=12cm}
\end{center}
\caption{Optimal (A) and non-optimal (B) reaction activity in the reconstructed metabolic network
of {\it E.\ coli} in glucose minimal medium (Materials and Methods). The pie charts show the fractions of active and inactive
reactions in the metabolic subsystems defined in the iJR904 database~\citep{Reed2003-s}. The color code is as follows:
active reactions (red), inactive reactions due to mass balance (black) and environmental constraints (blue),
inactive reactions due to the irreversibility (green) and cascading (yellow) mechanisms, and  conditionally
inactive reactions (orange), which are inactive reactions that can be active 
for other
growth-maximizing states under the same medium condition.
The optimal state shown in panel A 
was obtained by flux balance analysis (FBA, see Materials and Methods).  The network is constructed by drawing an arrow from one subsystem to another when there are at least 4 metabolites that can be produced by reactions in the first subsystem and consumed by reactions in the second. Larger pies represent subsystems with more reactions.}
\label{fig:global}
\end{figure}

To provide mechanistic insight into such behaviors, here we study
the metabolic system of single-cell organisms under
optimal {\it and} non-optimal conditions
in terms of 
the number of {\it active} reactions (those that are actually used).
We implement our study within a flux balance-based framework~\citep{bonarius1997,edwards1999b,price_rev_2004,varma1994,Segre2002,Price:2003gd}. Figure~\ref{fig:global} illustrates 
key aspects
of our 
analysis 
using the example of {\em Escherichia coli}. 
For any typical non-optimal state (Fig.~\ref{fig:global}B), all the reactions in the metabolic network are active, except for those that are necessarily inactive 
due either to
mass balance constraints or 
environmental conditions (e.g., nutrient limitation). In contrast, a large number of additional reactions are predicted to become inactive for any metabolic flux distribution maximizing the growth rate (Fig.~\ref{fig:global}A). This spontaneous reaction silencing effect, in which optimization causes massive reaction inactivation, is observed in all four organisms analyzed
in this study, {\it H.\ pylori}, {\it S.\ aureus}, {\it E.\ coli}, and {\it S.\ cerevisiae}, which have genomes
and metabolic networks of increasing size and complexity (Materials and Methods).
Our analysis reveals two mechanisms responsible for this effect:
(1)  {\em irreversibility} of a large number of reactions, which
under intracellular physiological conditions \citep{Feist:2007lr} is shared by more than 
62\% of all metabolic reactions in the organisms we analyze (Table 1 and Note \ref{Note1}); and
(2)  {\it cascade of inactivity} triggered by  
the irreversibility, which propagates
through the metabolic network 
due to
stoichiometric and 
flux 
balance constraints.
We provide
experimental evidence of this phenomenon and explore applications to data interpretation  
by analyzing intracellular 
flux and gene activity data available in the literature.

The drastic difference between optimal and non-optimal behavior is a general phenomenon 
 that we predict
not only for the maximization of growth, but also for the optimization of any typical objective function that is linear in metabolic fluxes, such as the production rate of a metabolic compound. Interestingly, we find that the resulting number of active reactions in optimal states is fairly constant across the four organisms analyzed, despite the significant differences in their biochemistry and in the number of available reactions. In glucose media, this number is $\sim 300$ and approaches the minimum required for growth, indicating that optimization tends to drive the metabolism surprisingly close
to the onset of cellular growth. 
This reduced number of active reactions is approximately the same for any typical objective function under the same growth conditions.

We suggest that these findings will have implications for 
the targeted improvement of cellular properties~\citep{Burgard-AP:2003eu}.
Recent work predicts that the knockout of specific enzyme-coding genes can enhance metabolic performance and even rescue otherwise nonviable strains~\citep{Motter:2008rm}.
The possibility of such knockouts bears on the issue of whether the inactivation of the corresponding enzyme-catalyzed reactions would bring the whole-cell metabolic  state  close to the target objective.
Thus, our identification of a cascading mechanism for inducing optimal reaction activity for arbitrary objective functions provides a natural set of candidate genetic interventions for the knockout-based enhancement of metabolic function~\citep{Motter:2008rm}.

\section*{Results}

\subsection*{Typical Non-optimal States}

We model  
cellular metabolism 
as a network of metabolites
connected through reaction and transport fluxes. 
The state of the system is represented by the  vector $\bv = (v_1,\ldots,v_N)^T$ of these fluxes, including the fluxes of  $n$ internal and transport reactions, as well as $n_\text{ex}$ exchange fluxes for modeling media conditions.
Under the constraints imposed by stoichiometry, reaction irreversibility, substrate availability,
and the assumption of steady-state conditions,  the state of the system is restricted to a 
{\em feasible solution space} $M \subseteq \R^N$ (Materials and Methods). 
Within this framework, we first  
consider the number of active reactions in a typical non-optimal state $\bv \in M$.

We can prove that, with the exception of the reactions that are inactive for all 
$\bv \in M$, all the metabolic reactions are active for almost all $\bv \in M$, 
i.e., for any typical state chosen randomly from $M$ (Text S1, Section 1).
Accordingly, the number $n_+(\bv )$ of active reactions in a typical non-optimal state 
is constant, i.e.,
\begin{equation}\label{eqn:subopt}
n_+(\bv) = n_+^\text{typ},
 \quad \text{for almost all } \bv \in M.
\end{equation}
The reactions that are inactive for all states are so either due to mass balance or environmental
conditions, and can be identified numerically using flux coupling~\citep{burgard2004} and flux 
variability analysis~\citep{Mahadevan:2003fk}.

\paragraph*{Mass balance.}
\ Part of the metabolic reactions are forced to be inactive solely due to mass balance, independently of the
medium conditions. 
For example, glutathione oxidoreductase in the {\it E.\ coli} reconstructed model involves oxidized glutathione,
but because there is no other metabolic reaction that can balance the flux of this metabolite, the reaction cannot
be active in any steady state. We characterize such reactions uniquely by a linear relationship between 
vectors of stoichiometric coefficients (Text S1, Section 2). Although these reactions are inactive in any steady state, some
of them may play a role in transient dynamics (e.g., after environmental changes)~\cite{Schuster:1991kx}, for which time-dependent analysis is required~\cite{Ingalls2003}.  
Others may be part of an incomplete pathway at an intermediate stage of the organism's evolution or, more likely, an artifact
of the incompleteness or stoichiometric inconsistencies of the reconstructed model.
Such inconsistencies have been identified in previous models~\cite{Poolman2006}, such as an earlier version of the model we use for {\it S. cerevisiae}~\cite{Gevorgyan:2008kx}.

\paragraph*{Environmental conditions.}
\ Other reactions are constrained to be inactive due to the constraints arising from 
the environmental conditions imposed by the medium. For example, all reactions in the
allantoin degradation pathway must be inactive for {\it E.\ coli} in 
media with no allantoin available,
since allantoin cannot be produced internally. Similarly, the reactions involved in
aerobic respiration are generally inactive for any state under anaerobic growth. 

\paragraph*{}

The results for the typical activity of each organism in glucose minimal media (Materials and Methods) are summarized in the
top\ bars of Fig.\ \ref{fig:nonzeros} and in Table~2.
The fraction of active reactions ranges from 50\%--82\%, while 9\%--23\% are inactive due to mass balance constraints 
and 9\%--26\% are inactive due to the environmental conditions.
Although the absolute number of active reactions tends to increase with the size of the metabolic network, the fraction
of active reactions appears to show the opposite tendency.
Figure\ \ref{fig:global}B shows that most of the subsystems of the {\it E.\ coli} metabolism 
are almost completely active, but a few have many inactive reactions. For example, 
due to the incompleteness of the network 
many reactions involving cofactors and 
prosthetic group biosynthesis cannot be used under
steady-state conditions
in any environment. 
In addition, many reactions in the alternate carbon metabolism, as well as many
transport and extracellular reactions, must be inactive in the absence of the corresponding substrates in
the glucose medium.

\subsection*{Growth-Maximizing States} 
\begin{figure}[ht]
\begin{center}
\epsfig{figure=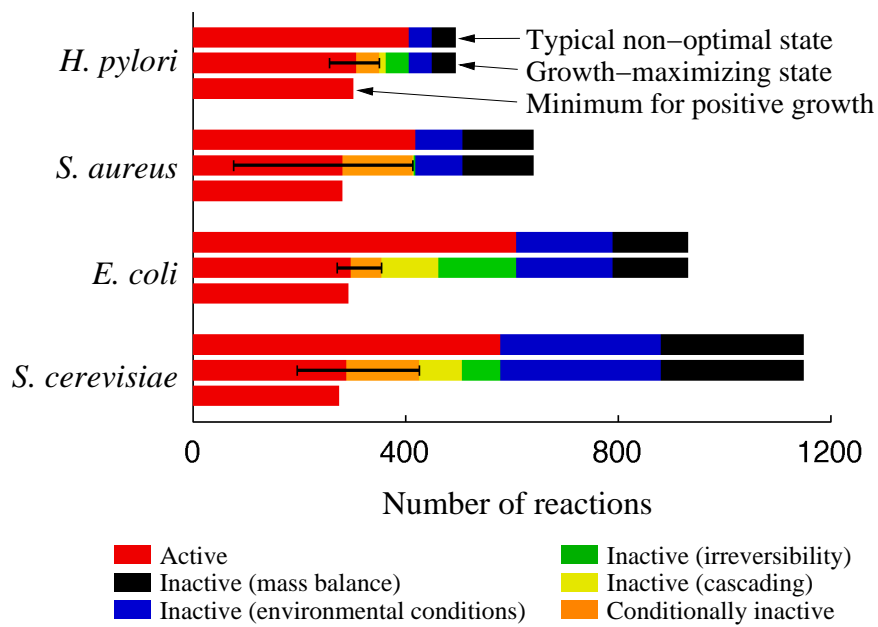,width=12cm}
\end{center}
\caption{Number of active and inactive reactions in the metabolic networks of {\it H.\ pylori}, {\it S.\ aureus}, {\it E.\ coli}, and {\it S.\ cerevisiae}. For each organism, the bars correspond to a typical non-optimal state (top),
a growth-maximizing state (middle), and a state with the minimum number of active reactions required for growth (bottom),
which was estimated using the algorithm described in Materials and Methods.
The error bar represents the upper and lower theoretical bounds, given by Eq.\ \eqref{naopt}, on the number of active reactions in any growth-maximizing state. 
The breakdown of inactive reactions and their color coding are the same as in Fig.~\ref{fig:global}.
All results are obtained using glucose minimal media (Materials and Methods) and are further detailed in Tables 2 and 3.}
\label{fig:nonzeros}
\end{figure}

We now turn to the maximization of growth rate, which is often hypothesized in flux balance-based
approaches and found to be consistent with observation 
in adaptive evolution experiments~\citep{edwards2000,edwards2001,Fong2004,pramanik1997}.
Performing numerical optimization in glucose minimal media (Materials and Methods), we find that the number of active reactions in such optimal states is reduced by 21\%--50\% compared
to typical non-optimal states, as indicated
in the middle bars of Fig.\ \ref{fig:nonzeros}.
Interestingly, the absolute number of active reactions under maximum growth is $\sim 300$ and appears to be
fairly independent of the organism and network size for the cases analyzed. 
We observe that the minimum number of reactions required merely to sustain positive growth\ \citep{Burgard:2001fk, Burgard:2001lr} is only a few reactions smaller than 
the number of reactions 
used in growth-maximizing states (bottom bars, Fig.\ \ref{fig:nonzeros}).
This implies that surprisingly small metabolic adjustment or genetic modification 
is sufficient for an optimally growing organism to stop growing 
completely, 
which
reveals a {\it robust-yet-subtle} tendency in cellular metabolism:   while the large number of  inactive reactions offers tremendous mutational and environmental robustness~\citep{Papp:2004dn}, the system is very sensitive if limited only to the set of reactions optimally active.  
The flip side of this prediction is that significant increase in growth can result from very few mutations, as observed recently in adaptive evolution experiments~\cite{Herring:2006fk}.

We now turn to mechanisms underlying the observed reaction silencing, which is spread over a wide range of metabolic subsystems (see Fig. 1 for {\it E.\ coli}).  The phenomenon is caused by growth maximization via reaction irreversibility and cascading of inactivity.

\begin{figure}[ht]
\begin{center}
\epsfig{figure=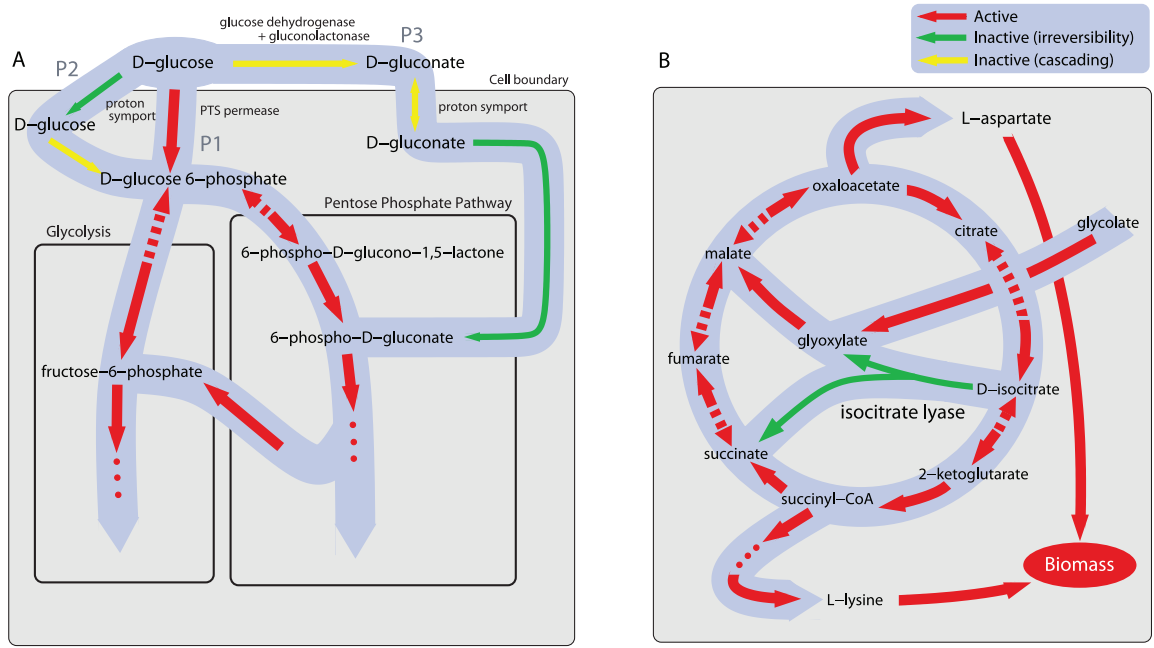, width=6in}
\end{center}
\caption{Portions of {\it E.\ coli} metabolic network under maximum growth condition.
(A) P1, P2, and P3 are alternative pathways for glucose transport and utilization.
The most efficient pathway, P1, is active under maximum growth 
in glucose minimal medium.
P2 and P3 are inactive, but if P1 is knocked out, P2 becomes active, and if both P1 and P2 
are knocked out, P3 becomes active. 
In both knockout scenarios, the growth is predicted to be suboptimal.
(B) Isocitrate lyase reaction in the pathway bypassing the tricarboxylic acid (TCA) cycle is predicted to be inactive under maximum growth due to its irreversibility.  If it were to operate in the opposite direction, it would serve as a {\em transverse} pathway which redirects 
metabolic flow to growth-limiting reactions, increasing the maximum biomass production rate slightly.
In both panels, single and double arrows are used to indicate irreversible and reversible reactions, respectively,
and colors indicate the behavior of the reactions under maximum growth: active (red), inactive due to the irreversibility (green), and inactive due to cascading (yellow).}
\label{fig:irr}
\end{figure}

\paragraph*{Irreversibility.}
\ We identify three different scenarios in which reaction irreversibility causes reaction inactivity under maximum growth.  The simplest case is when the reaction is part of a parallel pathway structure.
While stoichiometrically equivalent pathways lead to alternate optima~\citep{Mahadevan:2003fk},
``non-equivalent'' redundancy
can force irreversible reactions in less efficient pathways to be inactive. 
To illustrate this effect, 
we show in Fig.\ \ref{fig:irr}A three alternative pathways (P1, P2, and P3) for glucose transport and utilization in the {\it E.\ coli} metabolism.
Pathway P1 is active under maximum growth, while P2 and P3 are inactive because they are 
stoichiometrically less efficient for cellular growth.
Indeed, we computationally predict that knocking out P1 would make P2 active, but the growth rate would be
reduced by 2.5\%. Knocking out both P1 and P2 would make P3 active, but the growth rate would be reduced
by more than 10\%. 
Here, the irreversibility of P2 and P3 is essential.
For example, if P2 were reversible, the biomass production could be increased (by about 0.05\%) by making this pathway active in the opposite direction, which creates a metabolic cycle equivalent to a combination of the pyruvate kinase reaction and the transport of protons out of the cell.  The pyruvate kinase itself does not contribute to the increase in biomass production (it is inactive under maximum growth condition), but the cycle would provide a more efficient transport of protons to balance the influx of protons accompanying the ATP synthesis, which in turn would increase biomass production.

A different silencing 
scenario is identified when no clear parallel pathway structure is recognizable.
In this scenario there is a {\em transverse} pathway that, were it reversible, could be used to increase growth by redirecting metabolic flow from ``non-limiting'' pathways to those that limit the production of biomass precursors.
This includes transverse reactions that establish one-way communication between pathways that lead to different building blocks of the biomass (when one of them is more limiting than the others).
In the {\em E.~coli} model, for example, 
isocitrate lyase
in the glyoxylate bypass is predicted to be inactive under maximum growth, as shown in Fig.~\ref{fig:irr}B.  This prediction is consistent with the observation from growth experiments in glucose media~\cite{Fong2006}.  
Again, the irreversibility 
of the reaction (Note~\ref{Note2}) is essential for this argument because, if this constraint is hypothetically relaxed, we predict that the reaction becomes active in the opposite direction, which leads to a slight increase in the maximum growth rate (about 0.005\%).

A third scenario for the irreversibility mechanism arises when a transport reaction is irreversible because the corresponding substrate is absent in the 
medium.  
For example, since acetate, a possible carbon and energy source, is absent in the given medium, the corresponding transport reaction is irreversible; acetate can only go out of the 
cell (Note~\ref{Note3}).
For {\em E.~coli} under maximum growth, we computationally predict that this transport reaction is inactive.  
This indicates that {\em E.~coli} growing maximally in the given glucose medium wastes no acetate by excretion, which is consistent with experimental observation in glucose-limited culture at low dilution rate~\cite{AnkeKayser03012005}.
Our predictions in the previous section, in contrast, imply that acetate transport would be active in typical non-optimal states, suggesting that suboptimal growth may induce behavior that mimics acetate overflow metabolism.
More generally, we predict that 
a suboptimal cell will activate {\em more} transport reactions, and hence excrete {\em larger} number of metabolites than a growth-optimized cell.
This experimentally testable prediction can be further supported by our single-reaction knockout computations considered in a subsequent section (Experimental Evidence) to model suboptimal response to perturbation.

We interpret these inactivation mechanisms involving reaction irreversibility as a consequence of the linear
programming property that the set 
of optimal solutions $\mopt$ 
must be part of the boundary of $M$~\citep{best1985}.
As such, $\mopt$ is characterized by a set of \emph{binding} constraints, 
defined as inequality constraints (e.g., $v_i \le \beta_i$) satisfying two conditions: the equality holds ($v_i = \beta_i$) for all $\bv \in \mopt$ and \mopt\ is sensitive to changes in the constraints (changes in $\beta_i$).
In two dimensions, for example, $\mopt$ would be an edge of $M$, characterized by a single binding
constraint, or a corner of $M$, characterized by two binding constraints.
In general, at least $d-d_\text{opt}$ linearly independent constraints must be binding, 
where $d$ and $d_\text{opt}$ are the dimensions of $M$ and $\mopt$, respectively.
Since many metabolic reactions are subject to the irreversibility constraint ($v_i \ge 0$), 
this is expected to lead to many inactive reactions ($v_i = 0$). 
Indeed, 
by excluding the $k$ constraints that are not associated with reaction irreversibility (those for the ATP maintenance reaction and exchange fluxes), we obtain an upper bound on the number of active reactions $n_+(\bv)$:
\begin{equation}\label{nabnd}
n_+(\bv) \le n_+^\text{typ} - (d - d_\text{opt} - k).
\end{equation}

\paragraph*{Cascading.}
\ The remaining set of reactions that are inactive for all $\bv \in \mopt$ is due to cascading
of inactivity.
On one hand, if all the reactions that produce a metabolite are inactive, any reaction that consumes
this metabolite must be inactive.
On the other hand, if all the reactions that consume a metabolite are inactive, any reaction that
produces 
this metabolite must be inactive to avoid 
accumulation, as this
would violate the steady-state assumption.
Therefore,
the inactivity caused by the irreversibility mechanism triggers a cascade of inactivity both
in the forward and backward directions along the metabolic network. 
In general, there are many different sets of reactions that, when inactivated, can create the same cascading effect,
thus 
providing flexibility in potential 
applications of this effect
to
the design of optimal strains~\citep{Motter:2008rm}.
The cascades in the growth-maximizing states, however, are spontaneous,
as opposed to 
those that
would be induced
in metabolic knockout applications~\citep{Motter:2008rm} or in reaction essentiality and robustness studies~\citep{Ghim, Lemke,Smart:2008pi}.
Different but related to the cascades of inactivity are the concepts of enzyme subsets~\cite{Pfeiffer:1999rw}, coupled reaction sets~\cite{burgard2004} and correlated reaction sets~\cite{Reed:2004fk}, which describe groups of reactions that operate together and are thus concurrently inactivated in cascades.

\paragraph*{Conditional inactivity.}
\ While the irreversibility and cascading mechanisms cause the inactivity of many reactions for all $\bv \in \mopt$, the inactivity of other reactions can depend on the specific growth-maximizing state, whose non-uniqueness in a given environment has been evidenced both theoretically~\citep{Mahadevan:2003fk,Lee:2000lr,Reed:2004fk} and experimentally~\citep{Fong2005}.
To explore this dependence, we use
the duality principle of linear programming
problems~\citep{best1985}
to 
identify all the binding constraints generating the set 
of optimal
solutions $\mopt$ (Text S1, Section 3).
This characterization is then used to count the number $n_+^\text{opt}$ ($n_0^\text{opt}$) of reactions that are
active (inactive) for {\it all} $\bv \in \mopt$, leading to rigorous bounds for the number of active reactions $n_+(\bv)$:
\begin{equation}\label{naopt}
n_+^\text{opt} \le n_+(\bv) \le n - n_0^\text{opt}.
\end{equation}
Numerical values of the bounds under maximum growth are indicated by the error bars in Fig.\ \ref{fig:nonzeros}.
Note that the upper bounds are consistently smaller than $n_+^\text{typ}$ for typical non-optimal states, indicating that reaction silencing necessarily occurs for all growth-maximizing states.
For {\it E.\ coli}, these results are consistent with a previous study 
comparing reaction utilization under a range of different growth conditions~\citep{Reed:2004fk}.
They are also consistent with existing results for different {\it E.\ coli} metabolic models~\citep{Feist:2007lr,Henry:2006lr,Henry:2007lr} based on flux variability analysis~\citep{Mahadevan:2003fk}.
Furthermore, we can prove (Text S1, Section 3) that the distribution of $n_+(\bv)$ within the upper and lower bounds 
is singular in that the upper bound is attained for almost all optimal states: 
\begin{equation}\label{nadist}
n_+(\bv) = n - n_0^\text{opt}\quad \text{for almost all $\bv \in \mopt$}.
\end{equation}
Numerical simulations using standard simplex methods~\citep{glpk} 
actually result in much fewer active reactions, 
as shown in Fig.\ \ref{fig:nonzeros} (red middle bars),
because the algorithm finds solutions on the boundary of $\mopt$.
This behavior is expected, however, under the concurrent optimization of additional metabolic objectives, which generally tend to drive the flux distribution toward the boundary of $\mopt$.

\paragraph*{}

Figure \ref{fig:nonzeros} summarizes the
inactivity mechanisms
for the four organisms under maximum growth in glucose media (see also Fig.\ \ref{fig:global}),
showing the inactive reactions caused by the irreversibility (green) and cascading (yellow) mechanisms, as well as
those that are conditionally inactive (orange). 
Observe that in sharp contrast to the number of active reactions, which depends little on the size of the network,
the number of inactive reactions (either separated by mechanisms or lumped together) varies widely and shows non-trivial dependence on 
the organisms.

\subsection*{Typical Linear Objective Functions}
Although we have focused so far on maximizing the biomass production rate, 
the true nature of the fitness function driving evolution is far from clear\ \citep{Burgard2003,Schuetz:2007fk,Nolan2006, Gianchandani2008ld}.
Organisms 
under different conditions
may optimize different objective functions, such as ATP production or nutrient uptake, 
or not optimize a simple function at all.
In particular, some metabolic behaviors, such as the selection between respiration and fermentation in yeast, cannot
be explained by growth maximization~\cite{Schuster:2008kx}. Other behaviors may be systematically different in situations mimicking
natural environments~\cite{egli}. 
Moreover, various alternative target objectives
can be conceived and used in metabolic engineering applications.
We emphasize, however, that while specific numbers may differ 
in each case,
all the arguments leading to Eqs.\ \eqref{nabnd}--\eqref{nadist} are general and apply to any objective function that is linear in metabolic fluxes.
To obtain further insights, we now study the number of active reactions in organisms optimizing a typical linear objective function
by means of random uniform sampling.

We first note the fact (proved in Text S1, Section 4) 
that with probability one under uniform sampling, 
the optimal solution set
$\mopt$ consists of a single point, which must be a ``corner'' of $M$, 
termed an extreme point in the linear programming literature.
In this case, $d_\text{opt} = 0$, and Eq.~\eqref{nabnd} becomes
\begin{equation}\label{typ_opt}
n_+(\bv) \le n_+^\text{typ} - d + k.
\end{equation}
With the additional requirement that the organism show positive growth, we uniformly sample these extreme points, which represent all distinct optimal states for typical linear objective functions.

\begin{figure}[ht]
\begin{center}
\epsfig{figure=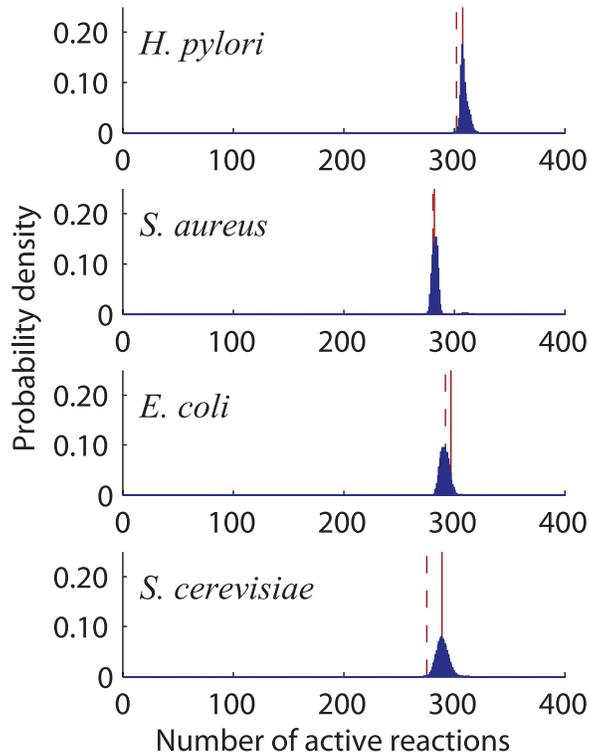,width=8cm}
\end{center}
\caption{Probability distribution of the number of active reactions in nonzero-growth states that optimize typical objective functions.
The red solid lines indicate the corresponding number in the growth-maximizing state 
of Fig.~\ref{fig:nonzeros} (middle bar, red), and the red dashed lines indicate 
our estimates of the minimum number of reactions required for the organism to grow (Materials and Methods).
[When the nonzero growth requirement is relaxed, a second sharp peak (not shown) arises, corresponding to a drop of $\sim 250$ in the number of active reactions caused by the inactivation of the biomass reaction.]}
\label{fig:sampling}
\end{figure}

We find that the number of active reactions in typical optimal states is narrowly distributed around that in growth-maximizing states, as shown in Fig.\ \ref{fig:sampling}.
This implies that various results on growth maximization extend to the optimization of typical objective functions.  
In particular, 
we see that a typical optimal state is 
surprisingly close to the onset of cellular growth (estimated and shown as dashed vertical lines in Fig.\ \ref{fig:sampling}). 
Despite the closeness, however, the organism maintains high {\em versatility}, which we define as the number of distinct functions that can be optimized under growth conditions.
To demonstrate this, consider the {\it H.\ pylori} model, which has 392 reactions that \emph{can} be active, among which at least 302 \emph{must} be active to sustain growth (Table 3).  
While only a few more than 302 active reactions are sufficient to optimize any objective function, the number of combinations for choosing them can be quite large.
For example, there are $\frac{(392-302)!}{(392-302-5)!5!} \approx 4 \times 10^{7}$ combinations for choosing, say, 5 extra reactions to be active.
Moreover, this number increases quickly with the network size: {\it S.\ cerevisiae}, for example, has less than 2.5 times more reactions than {\it H.\ pylori}, but about 500 times more combinations ($\frac{(579-275)!}{(579-275-5)!5!} \approx 2 \times 10^{10}$).

\subsection*{Experimental Evidence}
Our results help explain previous experimental observations. 
Analyzing the 22 intracellular fluxes 
determined
by Schmidt {\em et al.}~\cite{Schmidt1999} for the central metabolism of {\it E.\ coli} in both aerobic and anaerobic 
conditions, we find that  about 45\% of the fluxes are smaller than 10\% of the glucose uptake
rate (Table 4).  However, less than 19\% of the reversible fluxes and more than 60\% of the irreversible 
fluxes are found to be in this group (Fisher exact test, one-sided $p< 0.008$).  
For the 44 fluxes in the {\it S.\ cerevisiae} metabolism experimentally measured by Daran-Lapujade {\em et al.}~\cite{PascaleDaran-Lapujade03052004}, less than 8\% of the reversible fluxes and more than 42\% of the irreversible fluxes are zero (Table 5; Fisher exact test, one-sided $p< 10^{-7}$).
This higher probability
for reduced fluxes in irreversible reactions is consistent with 
our theory and simulation results (Table 6)
combined with the assumption
that the 
system operates close to an optimal state. 
For the {\it E.~coli} data, this assumption is justified by 
the work of \citet{Burgard2003}, where a framework for inferring metabolic objective functions was used
to show that
the organisms are mainly (but not completely) driven by the maximization of biomass production. 
The {\it S.~cerevisiae} data was also found to be consistent with the fluxes computed under the assumption of maximum growth~\cite{Papp:2004dn}.

Additional evidence for our results is derived from the inspection of 18 intracellular
fluxes experimentally determined by Emmerling {\em et al.}~\cite{Emmerling2002} for both wild-type {\it E.\ coli} and a 
gene-deficient strain not exposed to adaptive evolution. 
It has been
shown\ \citep{Segre2002} that while the wild-type fluxes can be approximately described by the 
optimization of biomass production, the genetically perturbed strain operates
sub-optimally. 
We consider the fluxes that are more than 10\% (of the glucose uptake rate) larger in the gene-deficient mutant
than in the wild-type strain. This group comprises less than 27\% of the reversible fluxes but more than 45\% of
the irreversible fluxes (Table 7; Fisher exact test, one-sided $p< 0.12$). This correlation indicates that irreversible fluxes tend to be larger in
non-optimal metabolic states, consistently with our predictions.

Altogether, our results offer an explanation for the temporary activation of latent pathways observed
in adaptive evolution experiments after environmental~\citep{Fong2005} or genetic perturbations~\citep{Fong2006}.
These initially inactive pathways are transiently activated after a perturbation, but subsequently inactivated again after adaptive evolution.
We hypothesize that {\em transient suboptimal states} are the leading cause of latent pathway activation.
Since we predict that a large number of reactions are inactive in optimal states but active in typical
non-optimal states, many reactions are expected to 
show temporary activation 
{\em if we assume} that the state following the perturbation is suboptimal and both the pre-perturbation and post-adaptation states are near optimality. 
To demonstrate this computationally for the {\it E.\ coli} model, we consider the idealized scenario where the perturbation to the growth-maximizing wild type is caused by a reaction knockout and the initial response of the metabolic network---before the perturbed strain evolves to a new growth-maximizing state---is well approximated by the method of minimization of metabolic adjustment (MOMA)~\citep{Segre2002}.
MOMA assumes that the perturbed organisms minimize the square-sum deviation of its flux distribution from the wild-type distribution (under the constraints imposed by the perturbation).

\begin{figure}[ht]
\begin{center}
\hspace{-1.5in}\epsfig{figure=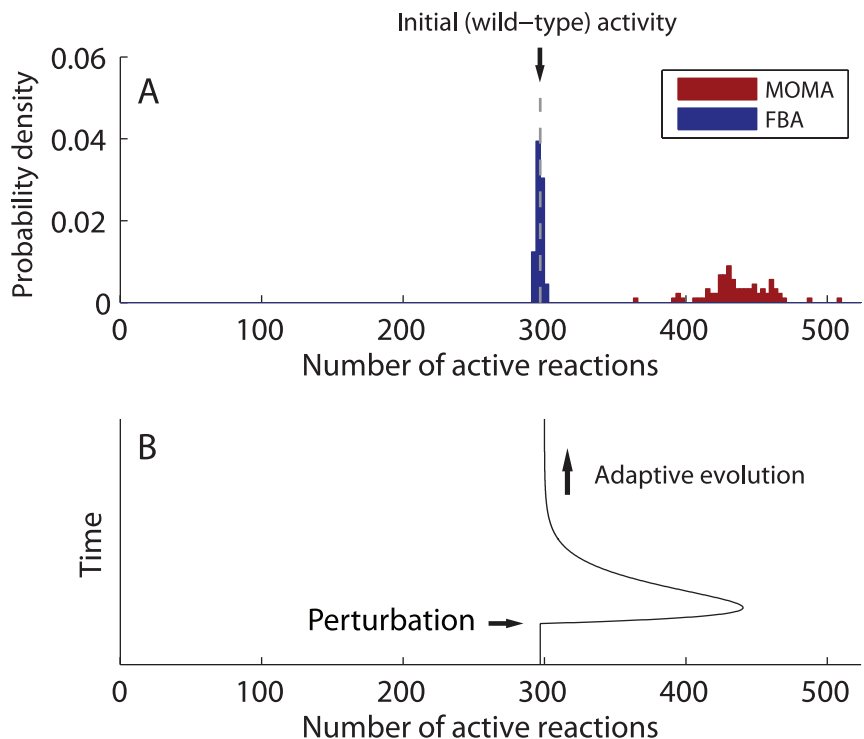}
\end{center}
\caption{Distribution of the number of active reactions in the {\it E.\ coli} metabolic network after a single-reaction knockout.  
(A) The initial response is predicted by the minimization of metabolic adjustment (MOMA) and the endpoint of adaptive evolution by the maximization of the growth rate (FBA), using the medium defined in Materials and Methods and a commercial optimization software package~\cite{cplex}.  
We consider all 77 nonlethal single-reaction knockouts that change the flux distribution.
(B) Schematic illustration of the network reaction activity during the adaptive evolution after a small perturbation, indicating the temporary activation of many latent pathways.}
\label{fig:knockouts}
\end{figure}

Figure~\ref{fig:knockouts}A shows the distribution of the number of active reactions for single-reaction knockouts that alter the flux distribution but allow positive MOMA-predicted growth.
While the distribution is spread around 400--500 for the suboptimal states in the initial response, it is sharply peaked around 300 for the optimal states at the endpoint of the evolution, which is consistent with our results on random sampling of the extreme points (Fig.~\ref{fig:sampling}). 
We thus predict that the initial number of active reactions for the unperturbed wild-type strain (which is 297, as shown by a dashed vertical line 
in Fig.~\ref{fig:knockouts}A) typically increases to more than 400 following the perturbation, and then decays back to a number close to 300 after adaptive evolution in the given environment, as illustrated schematically in Fig.~\ref{fig:knockouts}B.  A neat implication of our analysis is that the active reactions in the early post-perturbation state includes {\it much more}
than just a superposition of the reactions that are active in the pre- and post-perturbation optimal states, thus
explaining the pronounced burst in gene expression changes observed to accompany media changes and gene 
knockouts~\citep{Fong2005,Fong2006}.
For example, for {\em E.~coli} in glucose minimal medium, temporary activation is predicted for the Entner-Doudoroff pathway after {\em pgi} knockout and for the glyoxylate bypass after {\em tpi} knockout, in agreement with recent flux measurements in adaptive evolution experiments~\citep{Fong2006}.

Another potential application of our results is to explain previous experimental evidence that antagonistic pleiotropy is important in adaptive evolution~\citep{Cooper:2000qv}, as they indicate that increasing fitness in a single environment requires inactivation of many reactions through regulation and mutation of associated genes, which is likely to cause a decrease of fitness in some other environments~\cite{Hillenmeyer2008}.

\section*{Discussion}

Combining computational and analytical means, we have uncovered the microscopic mechanisms giving rise to the phenomenon of spontaneous reaction silencing in single-cell organisms, in which optimization of a single metabolic objective, whether growth or any other, significantly reduces the number of active reactions to a number that appears to be quite insensitive to the size of the entire network.
Two mechanisms have been identified for the large-scale metabolic inactivation: reaction irreversibility and cascade of inactivity. 
In particular, the reaction irreversibility inactivates a pathway when the objective function could be enhanced by hypothetically reversing the metabolic flow through that pathway.  We have demonstrated that such pathways can be found among non-equivalent parallel pathways, transverse pathways connecting structures that lead to the synthesis of different biomass components, and pathways leading to metabolite excretion.
Although the irreversibility and cascading mechanisms do not require explicit maximization of efficiency,
massive reaction silencing is also expected for organisms optimizing biomass yield or other linear functions
(of metabolic fluxes)
normalized by
uptake rates~\cite{Schuster:2008kx}. 
Furthermore, while we have focused on minimal media, we expect the effect to be even more pronounced in richer media.
On one hand, a richer medium has fewer absent substrates,
which increases the number of active reactions in non-optimal states. On the other hand, a richer medium
allows the organism to utilize more efficient pathways that would not be available in a minimal
medium, possibly further reducing the number of active reactions in optimal states.

Our study carries implications for both natural and engineered processes. 
In the rational design of microbial enhancement, for example, one seeks
genetic modifications that can optimize the production of specific metabolic compounds,
which is a special case of the optimization problem we consider here
and akin to the problem of identifying optimal reaction
activity~\citep{Burgard-AP:2003eu,Motter:2008rm}. The identification of a reduced set of active reactions also provides a new approach for testing
the existence of global metabolic objectives and their consistency with hypothesized objective functions~\cite{Schuetz:2007fk}.
Such an approach is complementary to current approaches based on
coefficients of importance\ \citep{Burgard2003,Nolan2006} or putative objective reactions~\cite{Gianchandani2008ld} and is expected to provide novel insights
into goal-seeking dynamic concepts such as cybernetic modeling\ \citep{Ramkrishna1987}. 
Our study may also help model 
compromises between competing goals, such as growth and
robustness  
against environmental or genetic changes \citep{Fischer2005}.

In particular, our results open a new avenue for addressing the origin of mutational robustness~\cite{deVisser2003,Pal:2003fv,Wagner:2005fu,Harrison:2007qy,Borenstein:2006qy,DeLuna:2008rp}. Single-gene
deletion experiments on {\em E.~coli} and {\em S.~cerevisiae} have shown that only a small
fraction of their genes are essential for growth under standard laboratory conditions~\cite{Baba:2006ng,Joyce12012006,Giaever:2002db}. The number
of essential genes can be even smaller given that growth defect caused by a gene deletion may
be synthetically rescued by compensatory gene deletions~\cite{Motter:2008rm}, an effect
not accounted for in single-gene deletion experiments. Under fixed environmental conditions,
large part of this mutational robustness arises from the reactions
that are inactive under maximum growth, whose deletion is predicted to have no effect on the
growth rate~\citep{Papp:2004dn}. For example, for {\it E.\ coli} in glucose medium, we predict
that 638 out of the 931 reactions can be removed simultaneously while retaining the maximum
growth rate (Note~\ref{Note4}),
which is comparable to 686 computed in a minimal genome study in rich media~\citep{Pal2006}.
But what is the origin of all these non-essential genes?

A recent study on {\em S.~cerevisiae} has shown that the single deletion of almost any non-essential
gene leads to a growth defect in at least one stress condition~\cite{Hillenmeyer2008}, providing
substantive support for the long-standing hypothesis that mutational robustness is a byproduct
of environmental robustness~\citep{Harrison:2007qy} (at least if we assume that the tested conditions are representative of the natural
conditions under which the organisms evolved). 
An alternative explanation would be that in variable environments, which is a natural selective pressure likely to be more important
than considered in standard laboratory experiments, the apparently dispensable pathways may play a significant role in suboptimal states induced by environmental changes.  This alternative is based on the hypothesis that latent pathways provide intermediate states necessary to
facilitate adaptation, therefore conferring competitive advantage {\em even if the pathways are not
active in any single fixed environmental condition}. 
In light of our results, this hypothesis can be tested experimentally in medium-perturbation assays by measuring the change in growth or other phenotype caused by deleting the predicted latent pathways in advance to a medium change.

We conclude by 
calling attention to 
a limitation and strength of our results, which have been obtained using steady-state analysis.  Such analysis avoids complications introduced by 
unknown regulatory and kinetic parameters, but admittedly does not account for constraints that could be introduced by the latter.
Nevertheless, we have been able to draw robust conclusions about dynamical behaviors, such as the impact of perturbation 
and 
adaptive evolution 
on  
reaction activity.
Our methodology scales well for genome-wide studies and may prove instrumental 
for  
the identification of specific extreme pathways~\cite{Schilling2000sg, Papin2002ue} or elementary modes~\cite{Schuster1994ge, Schuster2000uf} governing the optimization of metabolic objectives.
Combined with recent studies on complex networks~\citep{Vazquez:2003ek,RekaAlbert11012005,Almaas:2005vn,kaneko2006lic,Batada:2006fy,Albert-LaszloBarabasi07262007,Weitz2007bc} and the concept of functional modularity~\cite{Hartwell1999}, our results are likely to lead to new understanding of the interplay between \emph{network activity} and \emph{biological function}.

\subsection*{Notes}
\begin{enumerate}
\item \label{Note1} In addition, under steady-state conditions in the media considered in this study,
more than 77\% of the reversible reactions  
become constrained to be irreversible, rendering a total of more than 92\% of all reactions ``effectively'' irreversible.
\item \label{Note2} This reaction is regarded in the biochemical literature as irreversible under physiological conditions in the cell, and is constrained as such in the modeling literature~\cite{Feist:2007lr,edwards2000,Reed2003-s,Duarte2004-s}.
\item \label{Note3} Similar effective irreversibility is at work for any transport or internal reaction that is a unique producer of one or more chemical compounds in the cell.
\item \label{Note4} For single-reaction knockouts, MOMA predicts that 662 out of the 931 deletion mutants grow at more than 99\% of the wild-type growth rate.  Among these 662 reactions, 95\% are predicted to be inactive under maximum growth.
\end{enumerate}

\section*{Materials and Methods}

\subsection*{Strains and media}
All the stoichiometric data for the {\it in silico} metabolic systems used in our study are available at \url{http://gcrg.ucsd.edu/In_Silico_Organisms}.
For {\it H.~pylori} 26695~\citep{Thiele2005-s}, we used a medium with unlimited amount of water and protons, and limited amount of  oxygen (5 mmol/g DW-h), L-alanine, D-alanine, L-arginine, L-histidine, L-isoleucine, L-leucine, L-methionine, L-valine, glucose, Iron (II and III), phosphate, sulfate, pimelate, and thiamine (20 mmol/g DW-h).
For {\it S.~aureus} N315~\citep{Becker2005-s}, we used a medium with limited amount of glucose, L-arginine, cytosine, and nicotinate (100 mmol/g DW-h), and unlimited amount of iron (II), protons, water, oxygen, phospate, sulfate, and thiamin.
For {\it E.~coli} K12 MG1655~\citep{Reed2003-s}, we used a medium with limited amount of glucose (10 mmol/g DW-h) and oxygen (20 mmol/g DW-h), and unlimited amount of carbon dioxide, iron (II), protons, water, potassium, sodium, ammonia, phospate, and sulfate.
For {\it S.~cerevisiae} S288C~\citep{Duarte2004-s}, we used a medium with limited amount of glucose (10 mmol/g DW-h), oxygen (20 mmol/g DW-h), and ammonia (100 mmol/g DW-h), and unlimited amount of water, protons, phosphate, carbon dioxide, potassium, and sulfate.
The flux through the ATP maintenance reaction was set to 7.6 mmol/g DW-h for {\it E.\ coli} and 1 mmol/g DW-h for {\it S.\ aureus} and {\it S.\ cerevisiae}.


\subsection*{Feasible solution space}
Under steady-state conditions, a cellular metabolic state
is represented by a solution of a homogeneous linear equation describing the mass balance condition,
\begin{equation}\label{eqn:mb}
 \mathbf{S} \mathbf{v}=\mathbf{0},
\end{equation}
where $\mathbf{S}$ is the $m \times N$ stoichiometric matrix and $\bv \in \R^N$ is the vector of metabolic fluxes. 
The components of $\bv = (v_1,\dots,v_N)^T$ include the fluxes of $n$ internal and transport reactions as well as 
$n_\text{ex}$ exchange fluxes, 
which model the transport of metabolites across the system boundary.
Constraints of the form $v_i \le \beta_i$ imposed on the exchange fluxes are used to define the
maximum uptake rates of substrates in the medium.
Additional constraints of the form $v_i \ge 0$ arise for the reactions that are irreversible.
Assuming that the cell's operation is mainly limited by the availability of substrates in the medium,
we impose no other constraints on the internal reaction fluxes, except for the ATP maintenance flux
for {\it S.\ aureus}, {\it E.\ coli}, and {\it S.\ cerevisiae} (see Strains and media section above). The set of all flux vectors $\bv$ satisfying
the above constraints defines the {\it feasible solution space} $M \subset \R^N$, representing the
capability of the metabolic network as a system.  


\subsection*{Maximizing growth and other linear objective functions}
The flux balance analysis (FBA)~\cite{bonarius1997,edwards1999b,price_rev_2004,varma1994,Price:2003gd} used in this study is based on the maximization of a metabolic objective function $\bc^T \bv$ within the feasible solution space $M$, which is formulated as
a linear programming problem:
\begin{equation}\label{lp}
\begin{alignedat}{2}
&\text{maximize: } & \quad & \bc^T \bv = \sum_{i=1}^N c_i v_i\\
&\text{subject to: } & &\bS \bv = \0, \quad \bv  \in \R^N,\\
& & &\alpha_i \le v_i \le \beta_i,\quad i=1,\ldots,N.
\end{alignedat}
\end{equation}
We set $\alpha_i = -\infty$ if $v_i$ is unbounded below and  $\beta_i = \infty$ if $v_i$ is unbounded above.
For a given objective function, 
we numerically determine an optimal flux
distribution for this problem using an implementation of the
simplex method~\citep{glpk}.
In the particular case of growth maximization, the objective
vector $\bc$ is taken to be parallel to the biomass flux, which is modeled as an
effective reaction that converts metabolites into biomass.


\subsection*{Finding minimum reaction set for nonzero growth}
To find a set of reactions from which none can be removed without forcing zero growth, we start with the set of all reactions and recursively reduce it until no further reduction is possible. 
At each recursive step, we first compute how much the maximum growth rate would be reduced if each reaction were removed from the set individually.
Then, we choose a reaction that causes the least change in the maximum growth rate, and remove it from the set.
We repeat this step until the maximum growth rate becomes zero.
The set of reactions we have just before we remove the last reaction is a desired minimal reaction set.
Note that, since the algorithm is not exhaustive, the number of reactions in this set is an upper bound and approximation for the minimum number of reactions required to sustain positive growth.

\section*{Acknowledgements}
The authors thank Linda J. Broadbelt for valuable discussions and for providing feedback
on the manuscript.  The authors also thank Jennifer L. Reed and Adam M. Feist for providing information on their {\em in silico} models.

\section*{Supporting Information}
Text S1: Mathematical Results

\clearpage


\clearpage

\clearpage

\section*{Tables}
\baselineskip12pt
\subsubsection*{Table 1: Reversibility of metabolic reactions in the reconstructed networks.}
\begin{tabular}{@{\vrule height 10.5pt depth4pt width0pt}lp{17mm}p{17mm}p{15mm}l}
\hline
& {\it H.~pylori} & {\it S.~aureus} & {\it E.~coli} & {\it S.~cerevisiae}\\
\hline
Total number of reactions [$n$] : \hspace{1cm} & 479 & 641 & 931 & 1149\\
\hspace{5mm}Reversible & 165 & 220 & 245 & 430\\
\hspace{5mm}Irreversible & 314 & 421 & 686 & 719\\
\hline
\end{tabular}
\vspace{3mm}
\subsubsection*{Table 2: Metabolic reactions in typical non-optimal states of the simulated metabolisms.}
\begin{tabular}{@{\vrule height 10.5pt depth4pt  width0pt}lp{16mm}p{16mm}p{12mm}l}
\hline
& {\it H.~pylori} & {\it S.~aureus} & {\it E.~coli} & {\it S.~cerevisiae}\\
\hline
Total number of reactions [$n$] \hspace{1cm} & 479 & 641 & 931 & 1149\\
Inactive reactions: & 87 & 222 & 322 & 570\\
\hspace{5mm}Due to mass balance & 44 & 133 & 141 & 268\\
\hspace{5mm}Due to environmental conditions$^a$ \hspace{5mm} & 43 & 89 & 181 & 302\\
Active reactions [$n_+^\text{typ}$] & 392 & 419 & 609 & 579\\
\hline\\[-3mm]
\end{tabular}

\vspace{-1mm}
\noindent $^a$ {\footnotesize These reactions are inactive due to constraints arising from the availability of substrates in the media defined in Materials and Methods.}
\vspace{2mm}
\subsubsection*{Table 3: Metabolic reactions in maximum growth states of the simulated metabolisms.$^a$}
{\small
\begin{tabular}{@{\vrule height 10.5pt depth4pt  width0pt}lp{15mm}p{15mm}p{12mm}l}
\hline
& {\it H.~pylori} & {\it S.~aureus} & {\it E.~coli} & {\it S.~cerevisiae}\\
\hline
Active reactions under typical non-optimal states [$n_+^\text{typ}$]
& 392 & 419 & 609 & 579\\
Active reactions under maximum growth$^b$: & 308 & 282 & 297 & 289\\
\hspace{5mm}lower bound [$n_+^\text{opt}$] & 257 & 77 & 272 & 196 \\
\hspace{5mm}upper bound [$n - n_0^\text{opt}$] & 351 & 414 & 355 & 426 \\
Minimum number of active reactions for growth$^c$ \hspace{5mm} & 302 & 281 & 292$^e$ & 275 \\
\hline
Inactive reactions under maximum growth$^b$ [$n_0^\text{opt}$]: & 171 & 359 & 634 & 860\\
\hspace{5mm}Due to irreversibility & 29 & 3 & 147 & 72\\
\hspace{5mm}Due to cascading & 12 & 2 & 107 & 81 \\
\hspace{5mm}Due to mass balance & 44 & 133 & 141 & 268\\
\hspace{5mm}Due to environmental conditions \hspace{5mm} & 43 & 89 & 181 & 302\\
\hspace{5mm}Conditionally inactive$^d$ & 43 & 132 & 58 & 137 \\
\hline\\[-3mm]
\end{tabular}
}

\vspace{-1mm}
\noindent $^a$ {\footnotesize With respect to the minimal media defined in Materials and Methods.}\\
$^b$ {\footnotesize Based on a single optimal state found using an implementation of the simplex method~\citep{glpk}.}\\
$^c$ {\footnotesize Estimated using the algorithm described in Materials and Methods.}\\
$^d$ {\footnotesize Predicted to be inactive by the simplex method~\citep{glpk}, but can be active in some other growth-maximizing states.  Likewise, some of the reactions predicted to be active can be inactive in some other optimal states, but the number of such reactions is expected to be small since the simplex method finds a solution on the boundary of $M_\text{opt}$, which tends to have more inactive reactions than a typical optimal solution.}\\
$^e$ {\footnotesize The corresponding minimum number of  active reactions for maximum growth is 293.}

\subsubsection*{Table 4: Experimentally determined fluxes of intracellular reactions involved in the glycolysis, pentose phosphate pathway, 
TCA cycle, and amino acid biosynthesis of \textbf{\em E.~coli} K12 MG1655 under aerobic and anaerobic conditions\ \citep{Schmidt1999}.}{\small
\begin{tabular}{@{\vrule height 10.5pt depth4pt  width0pt}lcccc}
\hline
& \multicolumn{2}{c}{Aerobic} & \multicolumn{2}{c}{Anaerobic} \\
& Reversible & Irreversible & Reversible & Irreversible \\
\hline
Number of fluxes & 8 & 14 & 8 & 14 \\
Number of fluxes $<$ 10\% of glucose uptake rate & 1 & 7 & 2 & 10\\
\hline
\end{tabular}
}
\vspace{8mm}
\subsubsection*{Table 5: Experimentally determined fluxes of intracellular reactions involved in the  glycolysis, metabolic steps around pyruvate, TCA cycle, glyoxylate cycle, gluconeogenesis, and pentose phosphate pathway of \textbf{\em S.\ cerevisiae} strain CEN.PK113Ð7D grown under glucose, maltose, ethanol, and 
acetate limitation\ \citep{PascaleDaran-Lapujade03052004}.}{\small
\begin{tabular}{@{\vrule height 10.5pt depth4pt  width0pt}lcccccccc}
\hline
& \multicolumn{2}{c}{Glucose} & \multicolumn{2}{c}{Maltose} &  \multicolumn{2}{c}{Ethanol} & \multicolumn{2}{c}{Acetate}\\
& Rev. & Irr. & Rev. & Irr. & Rev. & Irr. & Rev. & Irr.\\
\hline
Number of fluxes & 22 & 22 & 22 & 22 & 22 & 22 & 22 & 22\\
Number of zero fluxes  & 2 & 8 & 2 & 7 & 1 & 11 & 2 & 11 \\
Percentage of zero fluxes & 9.1\% & 36.4\% & 9.1\% & 31.8\% & 4.5\% & 50.0\% & 9.1\% & 50.0\% \\
\hline
\end{tabular}
}
\vspace{8mm}
\subsubsection*{Table 6: Fraction of inactive reactions in the simulated metabolism of {\em E.~coli} and {\em S.~cerevisiae} under maximum growth condition.$^a$}
{\small
\begin{tabular}{@{\vrule height 10.5pt depth4pt  width0pt}lcccc}
\hline
& \multicolumn{2}{c}{\it E. coli} & \multicolumn{2}{c}{\it S. cerevisiae}\\
& Reversible & Irreversible & Reversible & Irreversible \\
\hline
Number of reactions & 245 & 686 & 430 & 719\\
Number of inactive reactions & 139 & 495 & 301 & 559\\
Percentage of inactive reactions & 56.7\% & 72.2\% & 70.0\% & 77.7\%\\
\hline\\[-3mm]
\end{tabular}
}

\noindent $^a$ {\footnotesize 
Same states considered in Table 3.
}
\vspace{8mm}
\subsubsection*{Table 7: Experimentally determined fluxes of reversible and irreversible reactions of wild-type \textbf{\em E.~coli} JM101 versus its pyruvate kinase-deficient
mutant PB25\ \citep{Emmerling2002}.}
{\small
\begin{tabular}{@{\vrule height 10.5pt depth4pt  width0pt}lll}
\hline
& Reversible & Irreversible \\
\hline
Number of fluxes & 30 & 24 \\
Number of mutant fluxes that are larger$^a$ by $>$ 10\% of glucose uptake rate\hspace{5mm} & 8 & 11\\
\hline
\end{tabular}
}

{\footnotesize \noindent $^a$ Relative to the corresponding fluxes in the wild-type strain.}

\baselineskip15pt

\begin{flushright} \small 
\textit{Spontaneous reaction silencing in metabolic optimization}\\
T.\ Nishikawa, N.\ Gulbahce, A.\ E.\ Motter
\end{flushright}

\vspace{2mm}
\begin{center}
\large Supporting Information\\[1mm]
\Large\bf Text S1: Mathematical Results
\end{center}

\setcounter{page}{1}
\setcounter{equation}{0}

\section*{1.\quad Number of active reactions in typical steady states}

\noindent The mass balance constraints $\bS \bv = \0$ define the linear subspace $\text{Nul}\,\bS = \{ \bv \in \R^N \,|\, \bS \bv = \0 \}$ (the null space of $\bS$), which contains the feasible solution space $M$.
However, the set $M$ can possibly be smaller than $\text{Nul}\,\bS$ because of the additional constraints arising from the environmental conditions (the availability of substrates in the medium, reaction irreversibility, and cell maintenance requirements).
Therefore, $M$ may have smaller dimension than $\text{Nul}\,\bS$.
If we denote the dimension of $M$ by $d$, there exists a unique $d$-dimensional linear submanifold of $\R^N$ that contains $M$, which we denote by $L_M$.
We can then use the Lebesgue measure naturally defined on $L_M$ to make probabilistic statements, since we can define the probability of a subset $A \subseteq M$ as the Lebesgue measure of $A$ normalized by the Lebesgue measure of $M$.
In particular, we say that $v_i \neq 0$ for {\em almost all} $\bv \in M$ if the set $\{ \bv \in M \,|\, v_i = 0 \}$ has Lebesgue measure zero on $L_M$.
An interpretation of this is that $v_i \neq 0$ with probability one for an organism in a random state under given environmental conditions. 
Using this notion, we prove the following theorem on 
the reaction fluxes.
\begin{theorem}\label{thm2}
If $v_i \neq 0$ for some $\bv \in M$, then $v_i \neq 0$ for almost all $\bv \in M$.
\end{theorem}
\begin{proof}
Suppose that $v_i \neq 0$ for some $\bv \in M$. The set $L_i := \{ \bv \in L_M \,|\, v_i = 0 \}$ is a linear submanifold of $L_M$, so we have $\dim{L_i} \le \dim{L_M}$. 
If $\dim{L_i} = \dim{L_M}$, then we have $L_i = L_M \supseteq M$, implying that we have $v_i = 0$ for any $\bv \in M$, which violates the assumption.
Thus, we must have $\dim{L_i} < \dim{L_M}$, implying that $L_i$ has zero Lebesgue measure on $L_M$.
Since $M \subseteq L_M$, we have $M_i := \{ \bv \in M \,|\, v_i = 0 \} \subseteq \{ \bv \in L_M \,|\, v_i = 0 \} = L_i$, and thus $M_i$ also has Lebesgue measure zero.
Therefore, we have $v_i \neq 0$ for almost all $\bv \in M$.
\end{proof}

Theorem~\ref{thm2} implies that we can group the reactions and exchange fluxes into two categories: 
\begin{enumerate}
\item {\em Always inactive}: $v_i = 0$ for all $\bv \in M$, and
\item {\em Almost always active}: $v_i \neq 0$ for almost all $\bv \in M$.
\end{enumerate}
Consequently, the number $n_+(\bv)$ of active reactions satisfies
\begin{equation}
n_+(\bv) = n_+^\text{typ} := n - n_0^m - n_0^e \quad \text{for almost all } \bv \in M,
\end{equation}
where $n_0^m$ is the number of inactive reactions due to the mass balance constraints (characterized by Theorem~\ref{thm1}) and $n_0^e$ is the number of additional reactions in the category 1 above, which are due to the environmental conditions.
Combining this result with the finding that optimal states have fewer active reactions (see the main text), it follows that a typical state $\bv \in M$ is non-optimal.

\section*{2.\quad Inactive reactions due to mass balance constraints}

\noindent Let us define the stoichiometric coefficient vector of reaction $i$ to be the $i$th column of the stoichiometric matrix $\bS$.
We similarly define the stoichiometric coefficient vector of an exchange flux.
If the stochiometric vector of reaction $i$ can be written as a linear combination of the stoichiometric vector of reactions/exchange fluxes $i_1, i_2, \ldots, i_k$, we say that $i$ is a linear combination of $i_1, i_2, \ldots, i_k$.
We use this linear relationship to completely characterize the set of all reactions that are always inactive due to the mass balance constraints, regardless of any additionally imposed constraints, such as the availability of substrates in the medium, reaction irreversibility, cell maintenance requirements, and optimum growth condition.

\begin{theorem}\label{thm1}
Reaction $i$ is inactive for all $\bv$ satisfying $\bS \bv = \0$ if and only if it is not a linear combination of the other reactions and exchange fluxes.
\end{theorem}

\begin{proof}
We denote the stoichiometric coefficient vectors of reactions and exchange fluxes by $\bs_1, \ldots, \bs_N$.
The theorem is equivalent to saying that there exists $\bv$ satisfying both $\bS \bv = \0$ and $v_i \neq 0$ if and only if $\bs_i$ is a linear combination of $\bs_k$, $k=1,2,\ldots,N$, $k \neq i$.

To prove the forward direction in this statement, suppose that $v_i \neq 0$ in a state $\bv$ satisfying $\bS \bv = \0$.
By writing out the components of the equation $\bS \bv = \0$ and rearranging, we get
\begin{equation}
s_{ji} v_i = \sum_{k \neq i} (- v_k) s_{jk}, \quad j=1,\ldots,m.
\end{equation}
Since $v_i \neq 0$, we can divide this equation by $v_i$ to see that $\bs_i$ is a linear combination of $\bs_k$, $k \neq i$ with coefficients $c_k = - v_k/v_i$.

To prove the backward direction, suppose that $\bs_i = \sum_{k \neq i} c_k \bs_k$.
If we choose $\bv$ so that $v_k = c_k$ for $k\neq i$ and $v_i = -1$, then for each $j$, we have
\[ (\bS \bv)_j = \sum_k v_k s_{jk} 
= v_i s_{ji} + \sum_{k\neq i} v_k s_{jk}
= -s_{ji} + \sum_{k\neq i} c_k s_{jk}
= 0, \]
so $\bv$ satisfies $\bS \bv = \0$.
\end{proof}

\section*{3. \quad Number of active reactions in optimal states}

\noindent The linear programming problem for finding the flux distribution maximizing a linear objective function can be written in the matrix form:
\begin{equation}\label{primal}
\begin{alignedat}{2}
&\text{maximize: } & \quad & \bc^T \bv \\
&\text{subject to: } & &\bS \bv = \0,\;
\bA \bv \le \bb, \; \bv \in \R^N,
\end{alignedat}
\end{equation}
where $\bA$ and $\bb$ are defined as follows.
If the $i$th constraint is $v_j \le \beta_j$, the $i$th row of $\bA$ consists of all zeros except for the $j$th entry that is $1$, and $b_i = \beta_j$.  
If the $i$th constraint is $\alpha_j \le v_j$, the $i$th row of $\bA$ consists of all zeros except for the $j$th entry that is $-1$, and $b_i = - \alpha_j$.  
A constraint of the type $\alpha_j \le v_j \le \beta_j$ is broken into two separate constraints and represented in $\bA$ and $\bb$ as above. 
The inequality between vectors is interpreted as inequalities between the corresponding components, so if the rows of $\bA$ are denoted by $\ba_1^T,\ba_2^T,\ldots,\ba_K^T$ (where $\ba_i^T$ denotes the transpose of $\ba_i$), $\bA \bv \le \bb$ represents the set of $K$ constraints $\ba_i^T \bv \le b_i$, $i=1,\ldots,K$.  By defining the feasible solution space 
\begin{equation}
M := \{ \bv \in \R^N \,|\, \bS \bv = \0,\; \bA \bv \le \bb \},
\end{equation}
the problem can be compactly expressed as maximizing $\bc^T \bv$ in $M$.

The duality principle~(Best \& Ritter, 1985) 
expresses that any linear programming problem (primal problem) is associated with a complementary linear programming problem (dual problem), and the solutions of the two problems are intimately related.
The dual problem associated with problem~\eqref{primal} is
\begin{equation}\label{dual}
\begin{alignedat}{2}
&\text{minimize: } & \quad & \bb^T \bu_1 \\
&\text{subject to: } & & \bA^T \bu_1 + \bS^T \bu_2 = \bc, \;
\bu_1 \ge \0, \\
& & &\bu_1 \in \R^K, \; \bu_2 \in \R^m,
\end{alignedat}
\end{equation}
where $\{ \bu_1, \bu_2 \}$ is the dual variable.
A consequence of the Strong Duality Theorem~(Best \& Ritter, 1985) is that the primal and dual solutions are related via a well-known optimality condition: $\bv$ is optimal for problem~\eqref{primal} if and only if there exists $\{ \bu_1, \bu_2 \}$ such that
\begin{gather}
\bS \bv = \0,\; \bA \bv \le \bb,\label{opt1}\\
\bA^T \bu_1 + \bS^T \bu_2 = \bc,\; \bu_1 \ge \0,\label{opt2}\\
\bu_1^T (\bA \bv - \bb) = 0. \label{opt3}
\end{gather}
Note that each component of $\bu_1$ can be positive or zero, and we can use this information to find a set of reactions that are forced to be inactive under optimization, as follows.
For any given optimal solution $\bv_0$, Eq.~\eqref{opt3} is equivalent to $u_{1i} (\ba_i^T \bv_0 - b_i) = 0$, $i=1,\ldots, K,$ where $u_{1i}$ is the $i$th component of $\bu_1$.
Thus, if $u_{i1}>0$ for a given $i$, we have $\ba_i^T\bv_0 = b_i$, and we say that the constraint $\ba_i^T\bv \le b_i$ is {\em binding} at $\bv_0$.
In particular, if an irreversible reaction ($v_i \ge 0$) is associated with a positive dual variable ($u_{1i} > 0$), then the irreversibility constraint is binding, and the reaction is inactive ($v_i = 0$) at $\bv_0$.
In fact, we can say much more: we prove the following theorem stating that such a reaction is actually {\em required to be inactive for all possible optimal solutions} for a given objective function $\bc^T \bv$.
\begin{theorem}\label{thm:optimal}
Suppose $\{ \bu_1, \bu_2 \}$ is a dual solution corresponding to an optimal solution of problem~\eqref{primal}.
Then, the set $\mopt$ of all optimal solutions of \eqref{primal} can be written as
\begin{equation}\label{opt}
\mopt = \{ \bv \in M \,|\,
\ba_i^T \bv = b_i \text{ for all $i$ for which $u_{1i} > 0$} \},
\end{equation}
and hence every reaction associated with a positive dual component
is binding for all optimal solutions in $\mopt$.
\end{theorem}
\begin{proof}[Sketch of proof]
Let $\bv_0$ be the optimal solution associated with $\{ \bu_1, \bu_2 \}$ and let $Q$ denote the right hand side of \eqref{opt}.
Any $\bv \in Q$ is an optimal solution of \eqref{primal}, since straightforward verification shows that it satisfies (\ref{opt1}-\ref{opt3}) with the same dual solution $\{ \bu_1, \bu_2 \}$.
Thus, we have $Q \subseteq \mopt$.
Conversely, suppose that $\bv$ is an optimal solution of \eqref{primal}.
Then, $\bv$ can be shown to belong to $H$, which we define to be the hyperplane that is orthogonal to $\bc$ and contains $\bv_0$, i.e.,
\begin{equation}
H := \{ \bv \in \R^N \,|\, \bc^T (\bv - \bv_0) = 0 \}. 
\end{equation}
This, together with the fact that $\bv$ satisfies $\bS \bv = \0$ and $\bA \bv \le \bb$, from \eqref{opt1}, can be used to show that $\bv \in Q$.
Therefore, any optimal solution must belong to $Q$.
Putting both directions together, we have $Q = \mopt$.
\end{proof}

Thus, once we solve Eq.~\eqref{primal} numerically and obtain a \emph{single} pair of primal and dual solutions ($\bv_0$ and $\{ \bu_1, \bu_2 \}$), we can use the characterization of $\mopt$ given in Eq.~\eqref{opt} to identify all reactions that are required to be inactive (or active) for any optimal solutions.
To do this we solve the following auxiliary linear optimization problems for each $i=1,\ldots,N$:
\begin{equation}\label{aux}
\begin{alignedat}{2}
&\text{maximize/minimize: } & \quad & v_i \\ 
&\text{subject to: } & &\bS \bv = \0,\; \bA \bv \le \bb,\;
\ba_i^T\bv = b_i \text{ for all $i$ for which $u_{1i}>0$.}
\end{alignedat}
\end{equation}
If the maximum and minimum of $v_i$ are both zero, then the corresponding reaction is required to be inactive for all $\bv \in \mopt$.
If the minimum is positive or maximum is negative, then the reaction is required to be active.
Otherwise, the reaction may be active or inactive, depending on the choice of an optimal solution.
Thus, we obtain the numbers $n_+^\text{opt}$ and $n_0^\text{opt}$ of reactions that are required to be active and inactive, respectively, for all $\bv \in \mopt$.
The number of active reactions for any $\bv \in \mopt$ is then bounded as
\begin{equation}\label{naopt-s}
n_+^\text{opt} \le n_+(\bv) \le n - n_0^\text{opt}.
\end{equation}

The distribution of $n_+(\bv)$ within the bounds is singular: the upper bound in Eq.~\eqref{naopt-s} is attained for almost all $\bv \in \mopt$.
To see this, we apply Theorem~\ref{thm2} with $M$ replaced by $\mopt$.
This is justified since we can obtain $\mopt$ from $M$ by simply imposing additional equality constraints.
Therefore, if we set aside the $n_0^\text{opt}$ reactions that are required to be inactive 
(including 
$n_0^m$ and $n_0^e$ reactions that are inactive for all $\bv \in M$), all the other reactions are active for almost all $\bv \in \mopt$.
Consequently,
\begin{equation}\label{qi}
n_+(\bv) 
= n - n_0^\text{opt}\quad \text{for almost all } \bv \in \mopt.
\end{equation}

We can also use Theorem~\ref{thm:optimal} to further classify those inactive reactions caused by the optimization as due to two specific mechanisms:
\begin{enumerate}
\item {\bf Irreversibility.} The irreversibility constraint ($v_i \ge 0$) on a reaction can be binding ($v_i = 0$), which directly forces the reaction to be inactive for all optimal solutions.  Such inactive reactions are identified by checking the positivity of dual components ($u_{1i}$).
\item {\bf Cascading.} All other reactions that are required to be inactive for all $\bv \in \mopt$ are due to a cascade of inactivity triggered by the first mechanism, which propagates over the metabolic network via the stoichiometric and mass balance constraints.
\end{enumerate}
In general, a given solution of problem~\eqref{primal} can be associated with multiple dual solutions.
The set and the number of positive components in $\bu_1$ can depend on the choice of a dual solution, and therefore the categorization according to mechanism is generally not unique.
As an example, consider a metabolic network containing a chain of two simple irreversible reactions, $A \xrightarrow{v_1} B \xrightarrow{v_2} C$.
Since the two reactions are fully coupled via the mass balance constraint ($v_1 = v_2$ whenever $\bS \bv = \0$), 
we can show that different combinations of dual components are possible for a given optimal solution: (i) $u_{11}>0, u_{12}=0$; (ii) $u_{11}=0, u_{12}>0$; or (iii) $u_{11}>0, u_{12}>0$.
In each case, the set of reactions in the irreversibility category is different, and the number of such reactions are  different in case (iii).
This comes from the fact that the same result ($v_1 = v_2 = 0$) follows from forcing $v_1=0$ only, $v_2=0$ only, or both.
Thus, we can interpret the non-uniqueness of the categorization as the fact that different sets of triggering inactive reactions can create the same cascading effect on the reaction activity.

\section*{4.\quad Typical linear objective functions}

Since the feasible solution space $M$ is convex, its ``corner'' can be mathematically formulated as an {\em extreme point}, defined as a point $\bv \in M$ that cannot be written as $\bv = a \bx + b \by$ with $a+b=1$, $0<a<1$ and ${\bf x, y} \in M$ such that ${\bf x \neq y}$.
Intuition from the two-dimensional case (Fig.~\ref{fig:extreme}) suggests that for a typical choice of the objective vector $\bc$ such that Eq.~\eqref{primal} has a solution, the solution is unique and located at an extreme point of $M$.
\setcounter{figure}{0}
\renewcommand{\thefigure}{S\arabic{figure}}
\begin{figure}
\begin{center}
\epsfig{figure=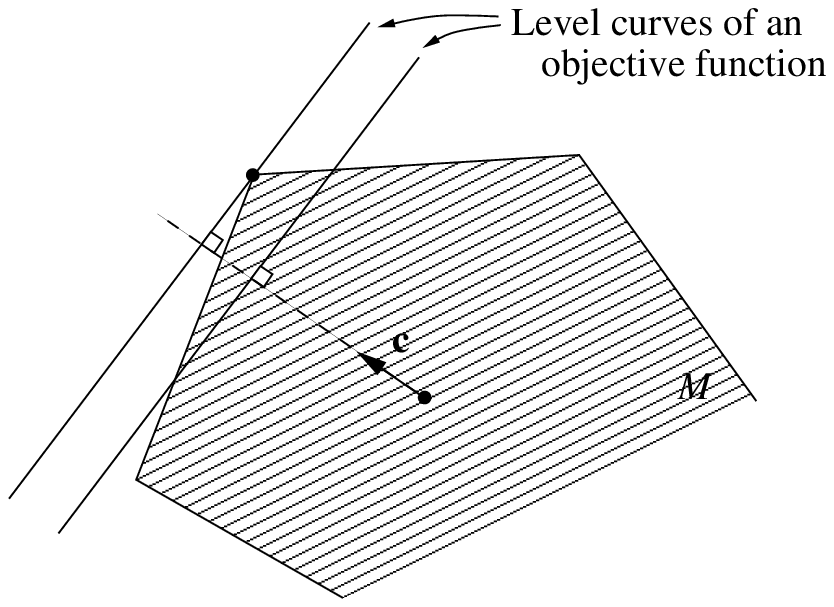,width=9cm}
\end{center}
\caption{Optimum is typically achieved at a single extreme point.  The only exception is when the objective vector $\bc$ is in the direction perpendicular to an edge, in which case all points on the edge are optimal.}
\label{fig:extreme}
\end{figure}
We prove here that this is indeed true in general, as long as the objective function is bounded on $M$, and hence an optimal solution exists.
\begin{theorem}\label{thm:corner}
Suppose that the set of objective vectors $B = \{\bc \in \R^N \,|\, \text{$\bc^T \bv$ is bounded on $M$}\}$ has positive Lebesgue measure.
Then, for almost all $\bc$ in $B$, there is a unique solution of Eq.~\eqref{primal}, and it is located at an extreme point of $M$.
\end{theorem}
\begin{proof}
For a given $\bc \in B$, the function $\bc^T \bv$ is bounded on $M$, so the solution set $\mopt = \mopt(\bc)$ of Eq.~\eqref{primal} consists of either a single point or multiple points.
Suppose $\mopt$ consists of a single point $\bv$ and it is not an extreme point.
By definition, it can be written as $\bv = a{\bf x}+b{\bf y}$ with $a+b=1$, $0<a<1$ and ${\bf x, y} \in M$ such that ${\bf x \neq y}$.
Since $\bv$ is the only solution of Eq.~\eqref{primal}, ${\bf x}$ and ${\bf y}$ must be suboptimal, and hence we have $\bc^T {\bf x} < \bc^T \bv$ and $\bc^T {\bf y} < \bc^T \bv$.
Then,
\begin{eqnarray*}
\bc^T {\bf y} &=& \bc^T (\bv - a{\bf x})/b\\
&=& (\bc^T \bv - a \bc^T {\bf x})/b\\
&>& (\bc^T \bv - a \bc^T \bv)/b\\
&=& \frac{1-a}{b} \, \bc^T \bv\\
&=& \bc^T \bv,
\end{eqnarray*}
and we have a contradiction with the fact that $\bv$ is an optimum.
Therefore, if $\mopt$ consists of a single point, it must be an extreme point of $M$.

We are left to show that the set of $\bc \in B$ for which $\mopt(\bc)$ consists of multiple points has Lebesgue measure zero.
By Theorem~\ref{thm:optimal}, for a given $\bc$, there exists a set of indices $I \subseteq \{1,\ldots,K\}$ such that $\mopt(\bc) = Q_I := \{ \bv \in M \,|\, \ba_i^T \bv = b_i \text{ for all $i \in I$} \}$, so
\begin{equation}\label{mopt}
\{ \bc \in \R^N \,|\, \mopt(\bc)\text{ contains multiple points} \}
\subseteq \bigcup_{I} \{\bc \in \R^N \,|\, Q_I = \mopt(\bc) \},
\end{equation}
where the union is taken over all $I \subseteq \{1,\ldots,K\}$ for which $Q_I$ contains multiple points.
If $\bc$ is in one of the sets in the union in Eq.~\eqref{mopt}, the set $Q_I$, being the set of all optimal solutions, is orthogonal to $\bc$.  Hence, $\bc$ is in $Q_I^\perp$, the orthogonal complement of $Q_I$ defined as the set of all vectors orthogonal to $Q_I$.
Therefore,
\begin{equation}\label{mopt2}
\{ \bc \in \R^N \,|\, \mopt(\bc)\text{ contains multiple points} \}
\subseteq \bigcup_{I} Q_I^\perp,
\end{equation}
Because $Q_I$ is convex, it contains multiple points if and only if its dimension is at least one, implying that each $Q_I^\perp$ in the union in Eq.~\eqref{mopt2} has dimension at most $N-1$, and hence has zero Lebesgue measure in $\R^N$.
Since there are only a finite number of possible choices for $I \subseteq \{1,\ldots,K\}$, the right hand side of Eq.~\eqref{mopt2} is a finite union of sets of Lebesgue measure zero.
Therefore, the left hand side also has Lebesgue measure zero.
\end{proof}


\vspace{5mm}
\noindent{\bf\Large Reference}

\noindent
Best MJ, Ritter K (1985)
{\it Linear Programming: Active Set Analysis and Computer Programs.}
Prentice-Hall, Engelwood Cliffs, New Jersey, USA


\baselineskip 15pt

\begin{thebibliography}{99}

\bibitem{Giaever:2002db}
Giaever G, Chu AM, Ni L, Connelly C, Riles L, et al. (2002)
Functional profiling of the {\it Saccharomyces cerevisiae} genome.
{\it Nature} {\bf 418:} 387--391

\bibitem{K.Kobayashi04152003}
Kobayashi K, Ehrlich SD, Albertini A, Amati G, Andersen KK, et al. (2003) 
\newblock {Essential {\em Bacillus subtilis} genes}.
\newblock {\em Proc Natl Acad Sci USA} {\bf 100:} 4678--4683

\bibitem{RosarioGil09012004}
Gil R, Silva FJ, Pereto J, Moya A (2004)
Determination of the core of a minimal bacterial gene set.
{\em Microbiol Mol Biol Rev} {\bf 68:} 518--537

\bibitem{Hashimoto2005}
Hashimoto M, Ichimura T, Mizoguchi H, Tanaka K, Fujimitsu K, et al. (2005)
Cell size and nucleoid organization of engineered {\em Escherichia coli} cells with a reduced genome.
{\em Mol Microbiol} {\bf 55:} 137--149

\bibitem{Baba:2006ng}
Baba T, Ara T, Hasegawa M, Takai Y, Okumura Y, et al. (2006)
Construction of {\em Escherichia coli} K-12 in-frame, single-gene knockout mutants: the Keio collection.
{\em Mol Syst Biol} {\bf 2:} 2006.0008

\bibitem{Joyce12012006}
Joyce AR, Reed JL, White A, Edwards R, Osterman A, et al. (2006)
Experimental and computational assessment of conditionally essential genes in {\em Escherichia coli}.
{\em J Bacteriol} {\bf 188:} 8259--8271

\bibitem[Burgard et~al(2001)]{Burgard:2001fk}
Burgard AP, Vaidyaraman S, Maranas CD (2001)
Minimal reaction sets for {\it Escherichia coli} metabolism under different growth requirements and uptake environments.
{\it Biotechnol Prog} {\bf 17:} 791--797

\bibitem[Burgard \& Maranas(2001)]{Burgard:2001lr}
Burgard AP, Maranas CD (2001)
Probing the performance limits of the {\it Escherichia coli} metabolic network subject to gene additions or deletions.
{\it Biotechnol Bioeng} {\bf 74:} 364--375

\bibitem[Mahadevan \& Schilling(2003)]{Mahadevan:2003fk}
Mahadevan R, Schilling C (2003)
The effects of alternate optimal solutions in constraint-based genome-scale metabolic models.
{\it Metab Eng} {\bf 5:} 264--276

\bibitem[Reed \& Palsson(2004)]{Reed:2004fk}
Reed JL, Palsson B{\O} (2004)
Genome-scale {\it in silico} models of {\it E. coli} have multiple equivalent phenotypic states: assessment of correlated reaction subsets that comprise network states.
{\it Genome Res} {\bf 14:} 1797--1805

\bibitem[P{\'a}l et~al(2006)]{Pal2006}
P{\'a}l C, Papp B, Lercher MJ, Csermely P, Oliver SG, et al. (2006)
Chance and necessity in the evolution of minimal metabolic networks.
{\it Nature} {\bf 440:} 667--670

\bibitem[Henry et~al(2006)]{Henry:2006lr}
Henry CS, Jankowski MD, Broadbelt LJ, Hatzimanikatis V (2006)
Genome-scale thermodynamic analysis of {\it Escherichia coli} metabolism.
{\it Biophys J} {\bf 90:} 1453--1461

\bibitem[Henry et~al(2007)]{Henry:2007lr}
Henry CS, Broadbelt LJ, Hatzimanikatis V (2007) 
Thermodynamics-based metabolic flux analysis.
{\it Biophys J} {\bf 92:} 1792--1805

\bibitem[Feist et~al(2007)]{Feist:2007lr}
Feist AM, Henry CS, Reed JL, Krummenacker M, Joyce AR, et al. (2007)
A genome-scale metabolic reconstruction for {\it Escherichia coli} K-12 MG1655 that accounts for 1260 ORFs and thermodynamic information.
{\it Mol Syst Biol} {\bf 3:} 121

\bibitem{Hillenmeyer2008}
Hillenmeyer ME, Fung E, Wildenhain J, Pierce SE, Hoon S, et al. (2008)
The chemical genomic portrait of yeast: uncovering a phenotype for all genes.
{\it Science} {\bf 320:} 362--365

\bibitem[Fong et~al(2005)]{Fong2005}
Fong SS, Joyce AR, Palsson B{\O} (2005)
Parallel adaptive evolution cultures of {\it Escherichia coli} lead to convergent growth phenotypes with different gene expression states.
{\it Genome Res} {\bf 15:} 1365--1372

\bibitem[Fong et~al(2006)]{Fong2006}
Fong SS, Nanchen A, Palsson B{\O}, Sauer U (2006)
Latent pathway activation and increased pathway capacity enable {\it Escherichia coli} adaptation to loss of key metabolic enzymes.
{\it J Biol Chem} {\bf 281:} 8024--8033

\bibitem[Varma \& Palsson(1994)]{varma1994}
Varma A, Palsson B{\O} (1994)
Metabolic flux balancing: basic concepts, scientific and practical use.
{\it Nat Biotechnol} {\bf 12:} 994--998

\bibitem[Bonarius et al(1997)]{bonarius1997}
Bonarius HPJ, Schmid G, Tramper J (1997)
Flux analysis of underdetermined metabolic networks: the quest for the missing constraints.
{\it Trends Biotechnol} {\bf 15:} 308--314

\bibitem[Edwards et~al(1999)]{edwards1999b}
Edwards JS, Ramakrishna R, Schilling CH, Palsson B{\O} (1999) 
Metabolic flux balance analysis.
In {\it Metabolic Engineering}, Lee SY, Papoutsakis ET (eds) pp 13--57. New York: CRC Press

\bibitem[Segr\`e et~al(2002)]{Segre2002}
Segr\`e D, Vitkup D, Church GM (2002)
Analysis of optimality in natural and perturbed metabolic networks.
{\it Proc Natl Acad Sci USA} {\bf 99:} 15112--15117

\bibitem{Price:2003gd}
Price ND, Papin JA, Schilling CH, Palsson B{\O} (2003)
Genome-scale microbial {\em in silico} models: the constraints-based approach.
{\em Trends Biotechnol} {\bf 21:} 162--169

\bibitem[Price et al(2004)]{price_rev_2004}
Price ND, Reed JL, Palsson B{\O} (2004)
Genome-scale models of microbial cells: evaluating the consequences of constraints.
{\it Nat Rev Microbiol} {\bf 2:} 886--897

\bibitem[Burgard et~al(2003)]{Burgard-AP:2003eu}
Burgard AP, Pharkya P, Maranas CD (2003)
Optknock: a bilevel programming framework for identifying gene knockout strategies for microbial strain optimization.
{\it Biotechnol Bioeng} {\bf 84:} 647--657

\bibitem[Motter et~al(2008)]{Motter:2008rm}
Motter AE, Gulbahce N, Almaas E, Barabasi A-L (2008)
Predicting synthetic rescues in metabolic networks.
{\it Mol Syst Biol} {\bf 4:} 168

\bibitem[Burgard et~al(2004)]{burgard2004}
Burgard AP, Nikolaev EV, Schilling CH, Maranas CD (2004)
Flux coupling analysis of genome-scale metabolic network reconstructions.
{\it Genome Res} {\bf 14:} 301--312

\bibitem{Schuster:1991kx}
Schuster S, Schuster R (1991)
Detecting strictly detailed balanced subnetworks in open chemical reaction networks.
{\it J Math Chem} {\bf 6:} 17--40

\bibitem{Ingalls2003}
Ingalls B, Sauro HM (2003)
Sensitivity analysis of stoichiometric networks: an extension of metabolic control analysis to non-steady state trajectories. 
{\it J Theor Biol} {\bf 222:} 23--36

\bibitem{Poolman2006}
Poolman MG, Bonde BK, Gevorgyan A, Patel HH, Fell DA (2006)
Challenges to be faced in the reconstruction of metabolic networks from public databases.
{\it Syst Biol (Stevenage)} {\bf 153:} 379--84

\bibitem{Gevorgyan:2008kx}
Gevorgyan A, Poolman MG, Fell DA (2008)
Detection of stoichiometric inconsistencies in biomolecular models.
{\it Bioinformatics} {\bf 24:} 2245--2251

\bibitem[Pramanik \& Keasling(1997)]{pramanik1997}
Pramanik J, Keasling JD (1997) 
Stoichiometric model of {\it Escherichia coli} metabolism: incorporation of growth-rate dependent biomass composition and mechanistic energy requirements.
{\it Biotechnol Bioeng} {\bf 56:} 398--421

\bibitem[Edwards \& Palsson(2000)]{edwards2000}
Edwards JS, Palsson B{\O} (2000)
The {\it Escherichia coli} MG1655 {\em in silico} metabolic genotype: its definition, characteristics, and capabilities.
{\it Proc Natl Acad Sci USA} {\bf 97:} 5528--5533

\bibitem[Edwards et~al(2001)]{edwards2001}
Edwards JS, Ibarra RU, Palsson B{\O} (2001)
{\em In silico} predictions of Escherichia coli metabolic capabilities are consistent with experimental data.
{\it Nat Biotechnol} {\bf 19:} 125--130

\bibitem[Fong \& Palsson(2004)]{Fong2004}
Fong SS, Palsson B{\O} (2004)
Metabolic gene-deletion strains of {\it Escherichia coli} evolve to computationally predicted growth phenotypes.
{\it Nat Genet} {\bf 36:} 1056--1058

\bibitem[Papp et~al(2004)]{Papp:2004dn}
Papp B, P{\'a}l C, Hurst LD (2004)
Metabolic network analysis of the causes and evolution of enzyme dispensability in yeast.
{\it Nature} {\bf 429:} 661--664

\bibitem{Herring:2006fk}
Herring CD, Raghunathan A, Honisch C, Patel T, Applebee MK, et al. (2006)
Comparative genome sequencing of {\em Escherichia coli} allows observation of bacterial evolution on a laboratory timescale.
{\em Nat Genet} {\bf 38:} 1406--1412

\bibitem{AnkeKayser03012005}
Kayser A, Weber J, Hecht V, Rinas U (2005)
Metabolic flux analysis of {\it Escherichia coli} in glucose-limited continuous culture. I. Growth-rate-dependent metabolic efficiency at steady state.
{\it Microbiology} {\bf 151:} 693--706

\bibitem[Best \& Ritter(1985)]{best1985}
Best MJ, Ritter K (1985)
{\it Linear Programming: Active Set Analysis and Computer Programs.}
Prentice-Hall, Engelwood Cliffs, New Jersey, USA

\bibitem[Lemke et~al(2004)]{Lemke}
Lemke N, Her\'edia F, Barcellos CK, dos Reis AN, Mombach JCM (2004)
Essentiality and damage in metabolic networks.
{\it Bioinformatics} {\bf 20:} 115--119

\bibitem[Ghim et~al(2005)]{Ghim}
Ghim CM, Goh K-I, Kahng B (2005)
Lethality and synthetic lethality in the genome-wide metabolic network of {\it Escherichia coli}.
{\it J Theor Biol} {\bf 237:} 401--411

\bibitem{Smart:2008pi}
Smart AG, Amaral LAN, Ottino JM (2008)
Cascading failure and robustness in metabolic networks.
{\it Proc Natl Acad Sci USA} {\bf 105:} 13223--13228

\bibitem{Pfeiffer:1999rw}
Pfeiffer T, Sanchez-Valdenebro I, Nuno J, Montero F, Schuster S (1999)
Metatool: for studying metabolic networks.
{\it Bioinformatics} {\bf 15:} 251--257

\bibitem[Lee et~al(2000)]{Lee:2000lr}
Lee S, Palakornkule C, Domach MM, Grossmann IE (2000)
Recursive MILP model for finding all the alternate optima in LP models for metabolic networks.
{\it Comput Chem Eng} {\bf 24:} 711--716

\bibitem[Makhorin(2001)]{glpk}
Makhorin A (2001)
GNU Linear Programming Kit (GLPK).
Available: \url{http://www.gnu.org/software/glpk/glpk.html}

\bibitem[Burgard \& Maranas(2003)]{Burgard2003}
Burgard AP, Maranas CD (2003)
Optimization-based framework for inferring and testing hypothesized metabolic objective functions.
{\it Biotechnol Bioeng} {\bf 82:} 670--677

\bibitem[Nolan et~al(2006)]{Nolan2006}
Nolan RP, Fenley AP, Lee K (2006)
Identification of distributed metabolic objectives in the hypermetabolic liver by flux and energy balance analysis.
{\it Metab Eng} {\bf 8:} 30--45

\bibitem[Schuetz et~al(2007)]{Schuetz:2007fk}
Schuetz R, Kuepfer L, Sauer U (2007)
Systematic evaluation of objective functions for predicting intracellular fluxes in {\it Escherichia coli}.
{\it Mol Syst Biol} {\bf 3:} 119

\bibitem{Gianchandani2008ld}
Gianchandani EP, Oberhardt MA, Burgard AP, Maranas CD, Papin JA (2008)
Predicting biological system objectives from internal state measurements. 
{\it BMC Bioinformatics} {\bf 9:} 43

\bibitem{Schuster:2008kx}
Schuster S, Pfeiffer T, Fell DA (2008)
Is maximization of molar yield in metabolic networks favoured by evolution?
{\it J Theor Biol} {\bf 252:} 497--504

\bibitem{egli}
Franchini AG, Egli T (2006) 
Global gene expression in {\it Escherichia coli} K-12 during short-term and long-term adaptation to glucose-limited continuous culture conditions. 
{\it Microbiology} {\bf 152:} 2111--2127

\bibitem[Schmidt et~al(1999)]{Schmidt1999}
Schmidt K, Nielsen J, Villadsen J (1999)
Quantitative analysis of metabolic fluxes in {\it Escherichia coli}, using two-dimensional NMR spectroscopy and complete isotopomer models.
{\it J Biotechnol} {\bf 71:} 175--190

\bibitem[Daran-Lapujade et~al(2004)]{PascaleDaran-Lapujade03052004}
Daran-Lapujade P, Jansen MLA, Daran JM, van Gulik W, de~Winde JH, et al. (2004)
Role of transcriptional regulation in controlling fluxes in central carbon metabolism of {\it Saccharomyces cerevisiae}: a chemostat culture study.
{\it J Biol Chem} {\bf 279}: 9125--9138

\bibitem[Emmerling et~al(2002)]{Emmerling2002}
Emmerling M, Dauner M, Ponti A, Fiaux J, Hochuli M, et al. (2002)
Metabolic flux responses to pyruvate kinase knockout in {\it Escherichia coli}.
{\it J Bacteriol} {\bf 184:} 152--164

\bibitem[Cooper \& Lenski(2000)]{Cooper:2000qv}
Cooper VS, Lenski RE (2000)
The population genetics of ecological specialization in evolving {\it Escherichia coli} populations.
{\it Nature} {\bf 407:} 736--739

\bibitem[Ramkrishna et~al(1987)]{Ramkrishna1987}
Ramkrishna D, Kompala DS, Tsao GT (1987)
Are microbes optimal strategists?
{\it Biotechnol Prog} {\bf 3:} 121--126

\bibitem[Fischer \& Sauer(2005)]{Fischer2005}
Fischer E, Sauer U (2005)
Large-scale {\it in vivo} flux analysis shows rigidity and suboptimal performance of {\it Bacillus subtilis} metabolism.
{\it Nat Genet} {\bf 37:} 636--640

\bibitem{Pal:2003fv}
P{\'a}l C, Papp B, Hurst LD (2003)
\newblock Rate of evolution and gene dispensability.
\newblock {\em Nature} {\bf 421:} 496--497

\bibitem{deVisser2003}
de Visser JAGM, Hermisson J, Wagner GP, Meyers LA, Bagheri-Chaichian H, et al. (2003)
Perspective: evolution and detection of genetic robustness.
{\it Evolution} {\bf 57:} 1959--1972

\bibitem{Wagner:2005fu}
Wagner A (2005)
Distributed robustness versus redundancy as causes of mutational robustness.
{\it BioEssays} {\bf 27:} 176--188

\bibitem[Borenstein \& Ruppin(2006)]{Borenstein:2006qy}
Borenstein E, Ruppin E (2006)
Direct evolution of genetic robustness in microRNA.
{\it Proc Natl Acad Sci USA} {\bf 103:} 6593--6598

\bibitem[Harrison et~al(2007)]{Harrison:2007qy}
Harrison R, Papp B, P{\'a}l C, Oliver SG, Delneri D (2007)
Plasticity of genetic interactions in metabolic networks of yeast.
{\it Proc Natl Acad Sci USA} {\bf 104:} 2307--2312

\bibitem{DeLuna:2008rp}
DeLuna A, Vetsigian K, Shoresh N, Hegreness M, Colon-Gonzalez M, et al. (2008)
Exposing the fitness contribution of duplicated genes.
{\it Nat Genet} {\bf 40:} 676--681
 
\bibitem{Schilling2000sg}
Schilling CH, Letscher D, Palsson B{\O} (2000)
Theory for the systemic definition of metabolic pathways and their use in interpreting metabolic function from a pathway-oriented perspective. 
{\it J Theor Biol} {\bf 203:} 229--248

\bibitem{Papin2002ue}
Papin JA, Price ND, Palsson B{\O} (2002)
Extreme pathway lengths and reaction participation in genome-scale metabolic networks.  
{\it Genome Res} {\bf 12:} 1889--1900

\bibitem{Schuster1994ge}
Schuster S, Hilgetag C (1994)
On elementary flux modes in biochemical reaction systems at steady state. 
{\it J Biol Syst} {\bf 2:} 165--182

\bibitem{Schuster2000uf}
Schuster S, Fell DA, Dandekar T (2000)
A general definition of metabolic pathways useful for systematic organization and analysis of complex metabolic networks. 
{\it Nat Biotechnol} {\bf 18:} 326--332

\bibitem[Vazquez et~al(2003)]{Vazquez:2003ek}
Vazquez A, Flammini A, Maritan A, Vespignani A (2003)
Global protein function prediction from protein-protein interaction networks.
{\it Nat Biotechnol} {\bf 21:} 697--700

\bibitem[Albert(2005)]{RekaAlbert11012005}
Albert R (2005)
{Scale-free networks in cell biology}.
{\it J Cell Sci} {\bf 118:} 4947--4957

\bibitem[Almaas et~al(2005)]{Almaas:2005vn}
Almaas E, Oltvai ZN, Barab{\'a}si A-L (2005)
The activity reaction core and plasticity of metabolic networks.
{\it PLoS Comput Biol} {\bf 1:} e68

\bibitem{Batada:2006fy}
Batada NN, Reguly T, Breitkreutz A, Boucher L, Breitkreutz BJ, et al. (2006) 
Stratus not altocumulus: a new view of the yeast protein interaction network.
{\it PLoS Biol} {\bf 4:} e317

\bibitem[Kaneko(2006)]{kaneko2006lic}
Kaneko K (2006)
{\it {Life: An Introduction to Complex Systems Biology}}.
Springer-Verlag, Berlin Heidelberg, Germany

\bibitem[Barab{\'a}si(2007)]{Albert-LaszloBarabasi07262007}
Barab{\'a}si AL (2007)
{Network Medicine -- From obesity to the ``diseasome.''}
{\it N Engl J Med} {\bf 357:} 404--407

\bibitem{Weitz2007bc}
Weitz JS, Benfey PN, Wingreen NS (2007)
Evolution, interactions, and biological networks.
{\it PLoS Biol} {\bf 5:} e11

\bibitem{Hartwell1999}
Hartwell LH, Hopfield JJ, Leibler S, Murray AW (1999)
From molecular to modular cell biology.
{\em Nature} {\bf 402:} C47--C52

\bibitem[Reed et~al(2003)]{Reed2003-s}
Reed JL, Vo TD, Schilling CH, Palsson B{\O} (2003)
An expanded genome-scale model of {\it Escherichia coli} K-12 (iJR904 GSM/GPR).
{\it Genome Biol} {\bf 4:} R54

\bibitem[Duarte et~al(2004)]{Duarte2004-s}
Duarte NC, Herrg{\aa}rd MJ, Palsson B{\O} (2004)
Reconstruction and validation of {\it Saccharomyces cerevisiae} iND750, a fully compartmentalized genome-scale metabolic model.
{\it Genome Res} {\bf 14:} 1298--1309

\bibitem[Thiele et~al(2005)]{Thiele2005-s}
Thiele I, Vo TD, Price ND, Palsson B{\O} (2005)
Expanded metabolic reconstruction of {\it Helicobacter pylori} (iIT341 GSM/GPR): an {\it in silico} genome-scale characterization of single- and double-deletion mutants.
{\it J Bacteriol} {\bf 187:} 5818--5830

\bibitem[Becker \& Palsson(2005)]{Becker2005-s}
Becker SA, Palsson B{\O} (2005)
Genome-scale reconstruction of the metabolic network in {\it Staphylococcus aureus} N315: an initial draft to the two-dimensional annotation.
{\it BMC Microbiol} {\bf 5:} 8

\bibitem{cplex}
ILOG CPLEX (Version 10.2.0). Available: \texttt{http://www.ilog.com/products/cplex/}

\end{thebibliography}
\end{document}